\documentclass[aps,preprint,nofootinbib,numerical]{revtex4-1} 
\newtheorem{Assumption}{\em A.}
\newtheorem{theorem}{Theorem}

\newtheorem{algoritmo}{\bf Algorithm}
\newtheorem{Lema}{\em Lemma}

\usepackage{subcaption}
\usepackage{graphicx}
\usepackage{latexsym}
\usepackage{amsfonts}
\usepackage{amssymb}
\usepackage{stmaryrd}
\usepackage{color}
\usepackage{blindtext}
\usepackage{amsmath}
\usepackage{MnSymbol}
\usepackage{caption}
\usepackage{mdframed}
\usepackage{epstopdf}
\usepackage{accents}

\newcommand{\QED}{\Box}
\newcommand{\rw}{\rightarrow}
\newcommand{\Natural}{\mathbb{N}}
\newcommand{\Real}{\mathbb{R}}
\newcommand{\mbE}{\mathbb{E}}

\newcommand{\mN}{{\mathcal N}}

\newcommand{\mO}{{\mathcal O}}

\newcommand{\sfa}{{\sf a}}
\newcommand{\sfb}{{\sf b}}
\newcommand{\sfc}{{\sf c}}
\newcommand{\sfd}{{\sf d}}
\newcommand{\sfh}{{\sf h}}

\newcommand{\bbf}{\boldsymbol{f}}
\newcommand{\bg}{\boldsymbol{g}}

\newcommand{\bv}{\boldsymbol{v}}
\newcommand{\bw}{\boldsymbol{w}}
\newcommand{\bx}{\boldsymbol{x}}
\newcommand{\by}{\boldsymbol{y}}

\newcommand{\bbF}{\boldsymbol{F}}
\newcommand{\bI}{\boldsymbol{I}}

\newcommand{\bP}{\boldsymbol{P}}

\newcommand{\bX}{\boldsymbol{X}}

\newcommand{\btheta}{\boldsymbol{\theta}}

\newcommand{\approptoinn}[2]{\mathrel{\vcenter{
			\offinterlineskip\halign{\hfil$##$\cr
				#1\sim\cr\noalign{\kern2pt}#1\propto\cr\noalign{\kern-2pt}}}}}

\def\BibTeX{{\rm B\kern-.05em{\sc i\kern-.025em b}\kern-.08em
    T\kern-.1667em\lower.7ex\hbox{E}\kern-.125emX}}

\begin{document}


\title{A probabilistic scheme for joint parameter estimation and state prediction in complex dynamical systems}

\author{Sara P\'erez-Vieites}
\email{spvieites@tsc.uc3m.es}
\affiliation{Department of Signal Theory \& Communications, Universidad Carlos III de Madrid. Avenida de la Universidad 30, 28911 Legan\'es, Madrid (Spain). }

\author{In\'es P. Mari\~no}
\email{ines.perez@urjc.es}
\affiliation{Department of Biology and Geology, Physics and Inorganic Chemistry, Universidad Rey Juan Carlos. C/ Tulip\'an s/n, 28933 M\'ostoles, Madrid, (Spain);\\and\\Department of Women's Cancer, Institute for Women's Health, University College London, London WC1E 6BT (United Kingdom).}

\author{Joaqu\'{\i}n M\'{\i}guez}
\email{joaquin.miguez@uc3m.es}
\affiliation{Department of Signal Theory \& Communications, Universidad Carlos III de Madrid. Avenida de la Universidad 30, 28911 Legan\'es, Madrid (Spain).}

\date{\today}

\begin{abstract}
Many problems in physics demand the ability to calibrate the parameters and predict the time evolution of complex dynamical models using sequentially-collected data. Here we introduce a general methodology for the joint estimation of the static parameters and the forecasting of the state variables of nonlinear stochastic dynamical models. The proposed scheme is essentially probabilistic. It aims at recursively computing the sequence of joint posterior probability distributions of the unknown model parameters and its (time varying) state variables conditional on the available observations. The new framework combines two layers of inference: in the first layer, a grid-based scheme is used to approximate the posterior probability distribution of the fixed parameters; in the  second layer, filtering (or {\em data assimilation}) techniques are employed to track and predict different conditional probability distributions of the state variables. Various types of procedures (deterministic grids, Monte Carlo, Gaussian filters, etc.) can be plugged into both layers, leading to a wealth of algorithms. For this reason, we refer to the proposed methodology as {\em nested hybrid filtering}. In this paper we specifically explore the combination of Monte Carlo and quasi Monte Carlo (deterministic) approximations in the first layer with Gaussian filtering methods in the second layer, but other approaches fit naturally within the new framework. We prove a general convergence result for a class of procedures that use sequential Monte Carlo in the first layer. Then, we turn to an illustrative numerical example. In particular, we apply and compare different implementations of the methodology to the tracking of the state, and the estimation of the fixed parameters, of a stochastic two-scale Lorenz 96 system. This model is commonly used to assess data assimilation procedures in meteorology. We show estimation and forecasting results, obtained with a desktop computer, for up to 5,000 dynamic state variables.  
\end{abstract}


\maketitle

\section{Introduction} \label{sIntro}

A common feature to many problems in some of the most active fields of science is the need to calibrate (i.e., estimate the parameters) and then forecast the time evolution of high-dimensional dynamical systems using sequentially-collected observations. One can find obvious examples in meteorology, where current models for global weather forecasting involve the tracking of millions of time-varying state variables \cite{Clayton13}, as well as in oceanography \cite{VanLeeuwen03} or in climate modelling \cite{Dee11}. This problem is not constrained to geophysics, though. In biochemistry and ecology it is often necessary to forecast the evolution of populations of interacting species (typically animal and/or vegetal species in ecology and different types of reacting molecules in biochemistry), which usually involves the estimation of the parameters that govern the interaction as well \cite{Golightly11}.

%
\subsection{State of the art}

Traditionally, model calibration (i.e., the estimation of the model static parameters) and the tracking and forecasting of the time-varying state variables have been addressed separately. The problem of tracking the state of the system using sequentially-collected observations is often termed {\em data assimilation} in geophysics, while it is referred to as {\em stochastic} or {\em Bayesian} filtering by researchers in computational statistics and applied probability. Carrying out both tasks jointly, parameter estimation and state forecasting, is a hard problem posing several practical and theoretical difficulties. 

Many procedures have been suggested over the years (see, e.g., \cite{Liu01b,Aksoy06,Evensen09,Carvalho10}, as well as \cite{Kantas15} for a survey), however they are subject to problems related to observability (i.e., ambiguities), lack of performance guarantees or prohibitive computational demands. Some of the most relevant techniques can be classified in one or more of the categories below.
\begin{itemize}
\item State augmentation methods with artificial dynamics: the state vector, which contains the dynamical variables that describe the physical system, is extended with any static unknown parameters (commonly reinterpreted as ``slowly changing'' dynamical variables) in the model \cite{Andrieu04,Kitagawa98,Liu01b,zhang2016joint}. Standard filtering (or data assimilation) techniques are then used in order to track and forecast the extended state vector. 
\item {\em Particle learning} techniques: for some models, the posterior probability distribution of the static parameters, conditional on the system state, can be computed in closed form and it depends only on a set of finite-dimensional statistics \cite{Storvik02,Djuric02,Carvalho10}. In a Monte Carlo setting, e.g., for particle filters, this means that the static parameters can be efficiently represented by sampling. Unfortunately, this approach is restricted to very specific models (an attempt to extend this idea to a more general setting can be found in \cite{Djuric04}). The term particle learning was coined in \cite{Carvalho10}, although the fundamental ideas were introduced earlier \cite{Storvik02,Djuric02}.
\item Classical importance resampling methods: several authors have studied the performance of classical sequential importance sampling for static parameters \cite{Papavasiliou05,Papavasiliou06,Olsson08}. Unfortunately, such algorithms tend to degenerate quickly over time unless certain conditions are met by the prior and posterior distributions \cite{Papavasiliou05,Papavasiliou06} or computationally-heavy interpolation schemes are adopted for the static parameters \cite{Olsson08}. 
\end{itemize}
		
Only in the last few years there have been advances leading to well-principled probabilistic methods that solve the joint problem numerically and supported by rigorous performance analyses \cite{Andrieu10,Chopin12,Crisan18bernoulli}. They aim at calculating the posterior probability distribution of {\em all the unknown variables and parameters} of the model. From the viewpoint of Bayesian analysis, these conditional, or {\em posterior}, distributions contain all the information relevant for the estimation task. From them, one can compute point estimates of the parameters and states but also quantify the estimation error. However, state-of-the-art methods for Bayesian parameter estimation and stochastic filtering are batch techniques, i.e., they process the whole set of available observations repeatedly in order to produce numerical solutions. For this reason, they are not well suited to problems where observations are collected sequentially and have to be processed as they arrive (or, simply, when the sequence of observations is too long). The popular particle Markov chain Monte Carlo (pMCMC) \cite{Andrieu10} and the sequential Monte Carlo square (SMC$^2$) \cite{Chopin12} schemes are examples of such batch methods. The nested particle filter (NPF) of \cite{Crisan18bernoulli} is a purely recursive Monte Carlo method, more suitable than pMCMC and SMC$^2$ when long sequences of observations have to be processed. However, this technique is still computationally prohibitive in high dimensional settings as it relies on two layers of intertwined Monte Carlo approximations.\\ 

While the schemes in \cite{Andrieu10,Chopin12,Crisan18bernoulli} fully rely on Monte Carlo approximations in order to approximate the posterior probability distribution of the parameters and the states, there is an alternative class of schemes, often coined recursive maximum likelihood (RML) methods \cite{Andrieu04,Andrieu12,Kantas15,Tadic10}, that enable the sequential processing of the observed data as they are collected but do not yield full posterior distributions of the unknowns. They only output point estimates instead. Therefore, it is not possible to quantify the uncertainty of the estimates or forecasts. Moreover, they are subject to various convergence (and complexity) issues, e.g., when the posterior probability distribution is multimodal, when it contains singularities or when the parameter likelihoods cannot be computed exactly.

In the physics literature, approximation schemes have been proposed that exploit the conditional dependences between the static parameters and the dynamic state variables, in a way that resembles the SMC$^2$ or NPF schemes. The authors of  \cite{santitissadeekorn2015two} introduce a two-stage filter that alternates the estimation of static parameters (conditional on a fixed state estimate) and the tracking of the dynamic variables (conditional on a fixed estimate of the static parameters). Another alternating scheme, that combines Monte Carlos estimators with ensemble Kalman filters in order to handle the static parameters and dynamic states, can be found in \cite{frei2012sequential}. 

In \cite{Ashley15}, an expectation-maximization (EM) algorithm is used to track a particle whose dynamics are governed by a hidden Markov model. The expectation step involves a (Monte Carlo based) forward-filtering, backward-smoothing step that is computationally heavy and prevents the online application of the method. The authors of \cite{Ye15} investigate a variational scheme (based in the Laplace integral approximation) for data assimilation (including state and parameter estimation) and illustrate it with applications to the Lorenz 63 and Lorenz 96 models in a low dimensional setting. The same task of data assimilation with parameter estimation is tackled in \cite{Ito16}. In this case, the estimation of the states and parameter is reduced to an optimization problem that can be solved via an adjoint method for the estimation of a Hessian matrix. The schemes in \cite{Ashley15}, \cite{Ye15} and \cite{Ito16} require to process the data in batches, rather than recursively, and hence they are not well suited for online implementations. A sequential method, based on variational Bayes techniques, that admits an online (recursive) implementation can be found in \cite{Vrettas15}. However, the latter contribution is limited to the estimation of the time-varying states and does not deal with unkown static parameters.
%
\subsection{Contribution}

In this paper we propose a general probabilistic scheme to perform the joint task of parameter estimation and state tracking and forecasting. The methodology is Bayesian, i.e., it aims at the computation of the posterior probability distribution of the unknowns given the available data. It involves two layers of estimators, one for the static parameters and another one for the time-varying state variables. It can be interpreted that the state estimators and predictors are {\em nested}, or inserted, within a main algorithm that tackles the estimation of the parameters. The estimation of the static parameters and the dynamic variables is carried out in a purely sequential and recursive manner. This property makes the proposed method well-suited for problems where long time series of data have to be handled. 
	
It can be shown that a particular case of the proposed scheme is the NPF of \cite{Crisan18bernoulli}, which relies on a sequential Monte Carlo sampler in the parameter space and bank of particle filters \cite{Gordon93,Doucet00} in the space of the dynamic variables. However, the key feature and advantage of the general scheme that we advocate here is the ability to combine different types of algorithms in the two layers of inference (parameters and dynamic variables). Any grid-based method (where the probability distribution of the static parameters is represented by a set of points in the parameter space) can be employed in the first layer, while the computationally-heavy particle filters in the second layer of the NPF can be replaced by simpler algorithms, easier to apply in practical problems. 

We have investigated the use of sequential Monte Carlo and quasi-Monte Carlo \cite{Gerber15} techniques in the parameter estimation layer. We note that the quasi-Monte Carlo scheme is a {\em deterministic} technique, although it formally resembles the Monte Carlo approach (hence the name). In the same vein, an unscented Kalman filter (UKF) can be utilized in the parameter estimation layer, although we have left this for future research. For the second layer, we have assessed two Gaussian filters, namely the extended Kalman filter (EKF) and the ensemble Kalman filter (EnKF). These two types of Gaussian filters have been well-studied in the geophysics literature and there are a number of numerical techniques to ease their practical implementation for large-scale systems (e.g., covariance inflation \cite{Anderson09, Lihong09} or localization \cite{Ott04,Houtekamer01,vossepoel2007parameter}). 
	
Because the flexibility to combine estimation techniques of different types within the same overall scheme is a key advantage of the proposed methodology, we refer to the resulting algorithms in general as {\em nested hybrid filters} (NHFs). Besides the numerical example described below, we provide a theoretical result on the asymptotic convergence of NHFs that use a sequential Monte Carlo scheme in the first layer (for the static parameters) and finite-variance estimators of the state variables in the second layer. Our analysis shows that the NHF can be biased if the filters in the second layer are (as it is the case in general with approximate Gaussian filters). However, it also ensures that the approximate posterior distribution of the parameters generated by the NHF, consisting of $N$ samples in the parameter space, converges to a well-defined limit distribution with rate $\mO\left(N^{-\frac{1}{2}}\right)$ under mild assumptions. 
	
To illustrate the performance of the methodology, we present the results of computer simulations with a stochastic two-scale Lorenz 96 model \cite{Arnold13} with underlying chaotic dynamics. In meteorology, the two-scale Lorenz 96 model is commonly used as a benchmark system for data assimilation \cite{VanLeeuwen10} and parameter estimation techniques \cite{Hakkarainen11} because it displays the basic physical features of atmospheric dynamics \cite{Arnold13} (e.g., convection and sensitivity to perturbations). We have implemented, and compared numerically, four NHFs that combine Monte Carlo, quasi-Monte Carlo, EKF and EnKF schemes in different ways. All the combinations that we have tried yield significant reductions of running times in comparison with the NPF for this model, without a significant loss of accuracy. We report simulation results for systems with up to 5,000 dynamical variables to track and forecast.
	

%
\subsection{Organization of the paper}
	
The rest of the paper is organized as follows. After a brief comment on notation, we describe in Section \ref{sDynamicalModelProblemStatement} the class of (stochastic) dynamical systems of interest. NHFs are introduced and explained in Section \ref{sNHFing}. The asymptotic convergence theorem is stated and discussed in Section \ref{sConvergence}. In Section \ref{sStochasticLorenz}, the stochastic Lorenz 96 model which is used in the simulations is described and then, some illustrative numerical results are presented in Section \ref{sNumericalResults}. Finally, Section \ref{sConclusions} is devoted to the conclusions.

%
\subsection{Notation}

We denote vectors and matrices by bold-face letters, either lower-case (for vectors) or upper-case (for matrices). Scalar magnitudes are denoted using regular-face letters. For example, $d \in \Natural$ and $x \in \Real$ are scalars, $\bx \in \Real^d$ is a vector and $\bX \in \Real^{d \times d}$ is a matrix. 
	
Most of the magnitudes of interest in this paper are random vectors (r.v.'s). If $\bx$ is a $d$-dimensional r.v. taking values in $\Real^d$, we use the generic notation $p(\bx)$ for its probability density function (pdf). This is an argument-wise notation. If we have two r.v.'s, $\bx$ and $\by$, we write $p(\bx)$ and $p(\by)$ for their respective pdf's, which are possibly different. In a similar vein, $p(\bx, \by)$ denotes the joint pdf of the two r.v.'s and $p(\bx | \by)$ denotes the conditional pdf of $\bx$ given $\by$. We find this simple notation convenient for the presentation of the model and methods and introduce a more specific terminology only for the analysis of convergence. We assume, for simplicity, that all random magnitudes can be described by pdf's with respect to the Lebesgue measure. Notation $\bx \sim p(\bx)$ is read as ``the r.v. $\bx$ is distributed according to the pdf $p(\bx)$''.
	
\section{Dynamical model and problem statement} \label{sDynamicalModelProblemStatement}

\subsection{State space models}  \label{sStateSpaceModel}

We are interested in systems that can be described by a multidimensional stochastic differential equation (SDE) of the form
\begin{equation} 
	\sfd \bx = \bbf(\bx, \btheta) \sfd t + \sigma \sfd \bw
	\label{eqStateCont}
\end{equation}
where $t$ denotes continuous time, $\bx(t) \in \Real^{d_x}$ is the $d_x$-dimensional system state, $\bbf:\Real^{d_x} \times \Real^{d_\theta} \rw \Real^{d_x}$ is a nonlinear function parametrized by a fixed vector of unknown parameters, $\btheta \in \Real^{d_\theta}$, $\sigma>0$ is a scale parameter that controls the intensity of the stochastic perturbation and $\bw(t)$ is a $d_x \times 1$ vector of independent standard Wiener processes. Very often, the underlying ordinary differential equation (ODE) $\dot \bx = \bbf(\bx,\btheta)$ describes some peculiar dynamics inherent to the system of interest (e.g., many of the systems of interest in geophysics are chaotic) and the addition of the perturbation $\bw(t)$ accounts for model errors or other sources of uncertainty.

Equation \eqref{eqStateCont} does not have a closed-form solution for a general nonlinear function $\bbf(\bx,\btheta)$ and, therefore, it has to be discretized for its numerical integration. A discretization scheme with fixed step-size $h>0$ yields, in general, a discrete-time stochastic dynamical system of the form
\begin{equation}
\bar \bx_k = \bar \bx_{k-1} + \bbF(\bar \bx_{k-1},\btheta,h,\sigma \bw_k),
\label{eqStateDiscrete}
\end{equation}
where $k \in \Natural$ denotes discrete time, $\bar \bx_k \simeq \bx(kh)$ is the system state at time $t=kh$ and $\bw_k$ is a r.v. of dimension $d_w \ge d_x$ that results from the integration of $d_w$ independent Wiener processes. Since the integral of a Wiener process over an interval of length $h$ is a Gaussian random variable with zero mean and variance $h$, the r.v. $\bw_k$ is also Gaussian, with mean $\boldsymbol{0}$ and covariance matrix $h\bI_{d_w}$, where $\bI_{d_w}$ is the $d_w \times d_w$ identity matrix. This was denoted as $\bw_k \sim \mN(\bw_k | {\bf 0},h\bI_{d_w})$. The function $\bbF$ depends on the choice of discretization scheme. The simplest one is the Euler-Maruyama method, which yields \cite{Gard88}
\begin{equation}
\bar\bx_k = \bar\bx_{k-1} + h\bbf(\bar\bx_{k-1},\btheta) + \sigma \bw_k,
\label{eqStateEuler}
\end{equation}
i.e., the noise is additive, with $d_w = d_x$, and $\bbF(\bar\bx_{k-1},\btheta,h,\sigma \bw_k)=h\bbf(\bar\bx_{k-1},\btheta) + \sigma \bw_k$. For a Runge-Kutta method of order $q$, as a more sophisticated example, $d_w =  qd_x$ and the function $\bbF$ results from applying $\bbf$ $q$ times, with a Gaussian perturbation passing through the nonlinearity at each of these intermediate steps. See \cite{Gard88} for details on various integration methods for SDEs. In the sequel, we work with the general Eq. \eqref{eqStateDiscrete}.

We assume that the system of Eq. \eqref{eqStateDiscrete} can be observed every $m$ discrete-time steps (i.e., every $mh$ continuous-time units). The $n$-th observation is a r.v. $\by_n \in \Real^{d_y}$ of dimension $d_y$ that we model as
\begin{equation} 
\by_n  = \bg(\bx_n, \btheta) + \bv_n,
\label{eqObservation}
\end{equation}
$n \in \Natural$, where $\bx_n = \bar \bx_{nm} \simeq \bx(nmh)$ is the system state at the time of the $n$-th observation (continuous time $t=nmh$), $\bg: \Real^{d_x} \times \Real^{d_\theta} \rightarrow \Real^{d_y} $ is a transformation that maps the state into the observation space and $\bv_n$ is a ${\bf 0}$-mean observational-noise vector with covariance matrix $\sigma_o^2 \bI_{d_y}$.

We can re-write the state Eq. \eqref{eqStateDiscrete} in the time scale of the observations (i.e., discrete-time $n$ rather than $k$) as
\begin{equation}
\bx_n = \bx_{n-1} + \bbF^m(\bx_{n-1},\btheta,h,\sigma \bw_n),
\label{eqStateDiscrete2}
\end{equation}
$n \in \Natural$, where the notation $\bbF^m$ indicates that Eq. \eqref{eqStateDiscrete} is applied $m$ consecutive times in order to move from $\bx_{n-1} = \bar\bx_{(n-1)m}$ to $\bx_n=\bar\bx_{nm}$. Note that, as a consequence, the noise vector $\bw_n$ in Eq. \eqref{eqStateDiscrete2} has dimension $d_w = mqd_x$.

\subsection{Probabilistic representation and problem statement} \label{sProblemStatement}

The state equation \eqref{eqStateDiscrete2} together with the observation equation \eqref{eqObservation} describe a (stochastic) state space model. The goal is to design methods for the recursive estimation of both the static parameters $\btheta$ and the states $\bx_n$, $n \in \Natural$. Note that the latter implies the estimation of the sequence $\bar\bx_k \simeq\bx(kh)$, $k \in \Natural$, i.e., the states between observations instants have to be estimated as well.

We adopt a Bayesian approach to this task. From this point of view, both the parameters $\btheta$ and the sequence of states $\bx_n$ are random and we aim at computing their respective probability distributions conditional on the available data, i.e., the sequence of observations $\by_n$. The problem is best described if we replace the functional representation of the state space model in Eqs. \eqref{eqStateDiscrete2} and \eqref{eqObservation} by an equivalent one in terms of pdf's \footnote{The pdf's in model \eqref{eqPrior_x}--\eqref{eqObservation_y} always exist if the functions $\bbf(\cdot,\cdot)$ and $\bg(\cdot,\cdot)$ are differentiable. There are problems for which the conditional probability distribution of $\bx_n$ conditional on $\bx_{n-1}$, which we may denote as $\tau_n$, may not have a density with respect to the Lebesgue measure and model \eqref{eqPrior_x}--\eqref{eqObservation_y} would not be well defined the way it is written. Even in that case, however, the proposed methodology could be applied using approximation filters in the second layer which do not depend on the differentiability of $\bbf$, such as particle filters or EnKF's.}. To be specific, the probabilistic representation consists of the following elements
\begin{eqnarray} 
\bx_0 &\sim& p(\bx_0) \label{eqPrior_x}\\
\btheta &\sim& p(\btheta) \label{eqPrior_theta}\\
\bx_n &\sim& p(\bx_n | \bx_{n-1},\btheta) \label{eqState_x}\\
\by_n &\sim& p(\by_n | \bx_n, \btheta) \label{eqObservation_y}
\end{eqnarray}
where $p(\bx_0)$ and $p(\btheta)$ are, respectively, the a priori pdf's of the system state and the parameter vector at time $n=0$ ($t=0$ as well), $p(\bx_n|\bx_{n-1}, \btheta)$ is the conditional pdf of $\bx_n$ given the state $\bx_{n-1}$ and the parameters in $\btheta $, and $p(\by_n|\bx_n, \btheta)$ is the conditional pdf of the observation given the state and the parameters. 

We note that:
\begin{itemize}
\item The priors $p(\bx_0)$ and $p(\btheta)$ can be understood as a probabilistic characterization of uncertainty regarding the system initial condition. If the initial condition $\bx_0$ were known, $p(\bx_0)$ could be replaced by a Dirac delta allocating probability 1 at that point. 
\item The pdf $p(\bx_n | \bx_{n-1}, \btheta)$ does not have, in general, a closed-form expression because of the nonlinearity $\bbF^m$. However, it is usually straightforward to simulate $\bx_n$ given $\bx_{n-1}$ and $\btheta$ using Eq. \eqref{eqStateDiscrete2} and this is sufficient for many methods to work.  
\item The observations are conditionally independent given the states and the parameters. If the observational noise $\bv_n$ is Gaussian, then $p(\by_n|\bx_n,\btheta) = \mN(\by_n | \bg(\bx_n,\btheta),\sigma_o^2\bI_{d_y})$.
\end{itemize}

From a Bayesian perspective, all the information relevant for the characterization of $\btheta$ and $\bx_n$ at discrete time $n$ (corresponding to $t=nmh$) is contained in the joint posterior pdf $p(\btheta,\bx_n|\by_{1:n})$, where $\by_{1:n} = \{ \by_1, \by_2, \ldots, \by_n \}$. The latter density cannot be computed exactly in general and the goal of this paper is to describe a class of flexible and efficient recursive methods for its approximation. 

We will show that one way to attain this goal is to tackle the approximation of the sequence of posterior pdf's of the parameters, $p(\btheta|\by_{1:n})$, $n \in \Natural$. This yields, in a natural way, approximations for $p(\btheta,\bx_n|\by_{1:n})$ and $p(\bx_n|\by_{1:n})$ for each $n$, as well as predictions for the densities of the intermediate states, $p(\bar\bx_{nm+k}|\by_{1:n})$, for $k = 1, \ldots, m-1$.

\section{Nested Hybrid Filtering}  \label{sNHFing}

\subsection{Importance sampling for parameter estimation} \label{ssIS}

In order to introduce the proposed scheme of nested hybrid filters, let us consider the approximation of the $n$-th posterior probability distribution of the parameters, with pdf $p(\btheta_n|\by_{1:n})$, using classical importance sampling \cite{Robert04}. In particular, let $q_n(\btheta)$ be an arbitrary proposal pdf for the parameter vector $\btheta$ and assume that $q_n(\btheta)>0$ whenever $p(\btheta|\by_{1:n})>0$.

Assume that the posterior at time $n-1$, $p(\btheta|\by_{1:n-1})$, is available. Then the posterior pdf at time $n$ can be expressed, via Bayes' theorem, as 
\begin{equation}
p(\btheta|\by_{1:n}) \propto p(\by_n|\btheta,\by_{1:n-1}) p(\btheta|\by_{1:n-1}), 
\label{eqBasic}
\end{equation}
where the proportionality constant, $p(\by_n|\by_{1:n-1})$, is independent of $\btheta$. Expression \eqref{eqBasic} enables the application of the importance sampling method to approximate integrals w.r.t. the posterior pdf $p(\btheta|\by_{1:n})$ (i.e., to approximate the statistics of this probability distribution). Specifically, if we
\begin{itemize}
\item draw $N$ independent and identically distributed (i.i.d.) samples from $q_n(\btheta)$, denoted $\btheta_n^i$, $i=1, \ldots, N$,
\item compute importance weights of the form 
$$
\tilde w_n^i = \frac{
	p(\by_n|\btheta_n^i,\by_{1:n-1}) p(\btheta_n^i|\by_{1:n-1})
}{
	q_n(\btheta_n^i)
},
$$
and normalize them to obtain
$$
w_n^i = \frac{
	\tilde w_n^i
}{
	\sum_{k=1}^N \tilde w_n^k
},	\quad i=1, \ldots, N,
$$
\end{itemize}
then it can be proved \cite{Robert04} that
\begin{equation}
\lim_{N\rw\infty} \sum_{i=1}^N w_n^i f(\btheta_n^i) = \int f(\btheta) p(\btheta|\by_{1:n}) \sfd \btheta \quad \mbox{almost surely (a.s.)}
\end{equation}
for any integrable function $f:\Real^{d_\theta}\rw\Real$ under mild regularity assumptions. In this way one could estimate the value of $\btheta$, e.g.,
$$
\btheta_n^N = \sum_{i=1}^N w_n^i \btheta_n^i \simeq  \int \btheta p(\btheta|\by_{1:n}) \sfd \btheta =: \mbE[\btheta|\by_{1:n}],
$$
where $ \mbE[\btheta|\by_{1:n}]$ denotes the expected value of $\btheta$ conditional on the observations $\by_{1:n}$. We could also estimate the mean square error (MSE) of this estimator, as
\begin{equation}
{\mbox{MSE}}^N_n = \sum_{i=1}^N w_n^i \| \btheta_n^i - \hat \btheta_n \|^2 \simeq \int \| \btheta_n^i - \mbE[\btheta|\by_{1:n}] \|^2 p(\btheta|\by_{1:n}) \sfd \btheta.
\end{equation}

The choice of $q_n(\btheta)$ is, of course, key to the complexity and the performance of importance sampling schemes. One particularly simple choice is $q_n(\btheta)=p(\btheta|\by_{1:n-1})$, which reduces the importance sampling algorithm to
\begin{enumerate}
\item drawing $N$ i.i.d. samples $\btheta_n^i$, $i=1, \ldots, N$, from $p(\btheta|\by_{1:n-1})$, and 
\item computing normalized importance weights
$
w_n^i \propto p(\by_n|\btheta_n^i,\by_{1:n-1}), \quad i=1, ..., N.
$
\end{enumerate}
Unfortunately, this method is not practical because
\begin{itemize}
	\item it is not possible to draw exactly from $p(\btheta | \by_{1:n})$, since this pdf is unknown, and
	\item the likelihood function $p(\by_n|\btheta_n^i,\by_{1:n-1})$ cannot be evaluated exactly either.
\end{itemize}
In the sequel we tackle the two issues above and, in doing so, obtain a general scheme for the approximation of the posterior distribution of the parameter vector $\btheta$ {\em and} the state vector $\bx_n$, i.e., the distribution with pdf $p(\btheta,\bx_n|\by_{1:n})$.


\subsection{Sequential Monte Carlo hybrid filter} \label{ssSMCfilter}

It is well known that the likelihood $u_n (\btheta) \coloneq p(\by_n | \btheta, \by_{1:n-1})$ can be approximated using filtering algorithms \cite{Andrieu10,Koblents15}. To be specific, function $u_n(\btheta)$ can be written as the integral
\begin{equation} \label{eqLikelihoodIntegral1}
u_n ( \btheta) = \int p(\by_n | \bx_n , \btheta) p( \bx_n | \btheta, \by_{1:n-1}) \sfd\btheta
\end{equation}
where, in turn, the predictive density $p(\bx_n | \btheta, \by_{1:n-1})$ is
\begin{equation} \label{eqPyn-1}
p( \bx_n | \btheta, \by_{1:n-1}) = \int p(\bx_n | \btheta, \bx_{n-1}) p( \bx_{n-1} | \btheta, \by_{1:n-1}) \sfd\bx_{n-1}
\end{equation}
and
\begin{equation} \label{eqPxn-1}
p( \bx_{n-1} | \btheta, \by_{1:n-1}) \propto p(\by_{n-1} | \btheta, \bx_{n-1}) p( \bx_{n-1} | \btheta, \by_{1:n-2}).
\end{equation}
Given a fixed parameter vector $\btheta$ and a prior pdf $p(\bx_0 | \btheta)$, the sequence of likelihoods $u_n(\btheta)$  can be computed by recursively applying Eqs. \eqref{eqLikelihoodIntegral1}, \eqref{eqPyn-1} and \eqref{eqPxn-1} for $n=1, 2, \ldots$.

Let us now assume that we are given a sequence of parameter vectors $\btheta_0, \ldots, \btheta_{k-1}, \btheta_k$ and we are interested in computing the likelihood of the last vector, $\btheta_k = \btheta'$. Following \cite{Crisan18bernoulli}, one can compute a sequence of \textit{approximate} likelihoods $\hat u_n (\btheta_n)$, $n=1, \ldots, k$, using the recursion
 \begin{eqnarray}
 \hat p(\bx_{n-1} | \btheta_{n-1}, \by_{1:n-1}) &\propto& p(\by_{n-1} | \btheta_{n-1}, \bx_{n-1})  \hat p(\bx_{n-1} | \btheta_{n-1}, \by_{1:n-2}) \label{eqRecur1} \\
 \hat p(\bx_n | \btheta_{n}, \by_{1:n-1}) &:=& \int p(\bx_n | \btheta_{n}, \bx_{n-1}) \hat p(\bx_{n-1} | \btheta_{n-1}, \by_{1:n-1}) \sfd \bx_{n-1}  \label{eqRecur2}
 \\
 \hat u_n(\btheta_{n}) &:=& \int p(\by_n | \btheta_{n}, \bx_n) \hat p(\bx_{n} | \btheta_n, \by_{1:n-1}) \sfd 
 \bx_{n} \label{eqRecur3}
 \end{eqnarray}
which starts with the initial density $\hat p(\bx_0|\btheta_0) := p(\bx_0 | \btheta_0)$. It can be proved, using the same type of continuity arguments in \cite{Crisan18bernoulli}, that the approximation error 
\begin{equation}
| u_k (\btheta') - \hat u_k(\btheta') |,
\label{eqError_k}
\end{equation}
can be kept bounded, for any $k$, provided some simple assumptions on the state space model and the sequence $\btheta_1, \ldots, \btheta_n$ are satisfied. Note that, in expression \eqref{eqError_k}, $u_k(\btheta')$ is the actual likelihood calculated by iterating \eqref{eqLikelihoodIntegral1}, \eqref{eqPyn-1} and \eqref{eqPxn-1} for $n=1, ..., k$, while $\hat u_k(\btheta')$ is the approximation computed using the sequence $\btheta_0, \ldots, \btheta_{k-1}, \btheta_k=\btheta'$ and recursion \eqref{eqRecur1}--\eqref{eqRecur3}.

The recursive approximation scheme for $\hat u_n(\btheta)$ can be combined with the ``naive" IS procedure of Section \ref{ssIS} to yield a general (and practical) method for the approximation of the sequence of \textit{a posteriori} probability distributions of the parameter vector $\btheta$, hereafter denoted as
$$
\mu_n (\sfd\btheta) := p(\btheta | \by_{1:n}) \sfd\btheta.
$$ 
We refer to the proposed scheme as a nested hybrid filter (NHF) and provide a detailed outline in Algorithm \ref{alNHF}.

		\begin{algoritmo} \label{alNHF}
	Nested hybrid filter (NHF).
	
	\textbf{Inputs}:
	\begin{itemize}
		\item[-] Number of Monte Carlo samples, $N$.
		\item[-] \textit{A priori} pdf's  $p(\btheta)$ and $p(\bx_0)$.
		\item[-] A Markov kernel $\kappa_N(\sfd\btheta |  \btheta')$ which, given $\btheta'\in D$, generates \textit{jittered} parameters $\btheta \in \Real^{d_\theta}$.
	\end{itemize}

	\textbf{Procedure}:
	\quad \begin{enumerate}	
		\item Initialization
		
		Draw $\btheta_0^{i}, i=1, \ldots,N$, i.i.d. samples from $\mu_0 (\sfd\btheta) = p(\btheta) \sfd\btheta$.
		
		\item Recursive step		
		\begin{enumerate}
			\item \label{enStep2a} For $i=1,\ldots,N$:
			\begin{enumerate}
				\item \label{enStep2a1} Draw $\bar{\btheta}_n^{i}$  from $\kappa_N (\sfd\btheta | \btheta_{n-1}^i)$.
				
				\item  \label{enStep2a2} Approximate $\hat p(\bx_n | \bar{\btheta}_n^i , \by_{1:n-1} )$ using a filtering algorithm.
				
				\item \label{enStep2a3} Use this approximation to compute the estimate
				\begin{equation}
				\hat{u}_n (\bar{\btheta}_n^i) = \int p(\by_n | \bar{\btheta}_n^i , \bx_n) \hat p(\bx_n | \bar{\btheta}_n^i , \by_{1:n-1} ) \sfd\bx_n 
				\end{equation}
				and let $w_n^i \propto \hat{u}_n(\bar{\btheta}_n^i)$ be the normalized weight of $\bar{\btheta}_n^i$.
			\end{enumerate}
			
			\item \label{enStep2b} Resample the discrete distribution 
			\begin{equation}
			\bar{\mu}_n^N (\sfd\btheta) = \sum_{i=1}^{N} w_n^i \delta_{\bar{\btheta}_n^i} ( \sfd\btheta)
			\end{equation}
			$N$ times with replacement in order to obtain the particle set $\{ \btheta_n^i\}_{i=1}^{N} $ and the approximate probability measure $\mu_n^N (\sfd\btheta) = \frac{1}{N} \sum_{i=1}^{N} \delta_{\btheta_n^i} (\sfd\btheta) $.
		\end{enumerate}
		
	\end{enumerate}
	
	\textbf{Outputs}: A set of particles $ \{ \btheta_n^i\}_{i=1}^{N} $ and a probability measure $\mu_n^N (\sfd\btheta)$.
		
\end{algoritmo}

Algorithm \ref{alNHF} is essentially a sequential Monte Carlo (SMC) method, often known as a particle filter \cite{Gordon93,Liu98,DelMoral04}. At each time step $n$, the output of the algorithm is an estimate of the posterior probability distribution $\mu_{n}(\sfd\btheta) = p (\btheta | \by_{1:n}) \sfd\btheta$. Specifically we construct the discrete and random probability measure
\begin{equation}
\mu_{n}^N (\sfd\btheta) = \frac{1}{N} \sum \delta_{{\btheta}_n^i} (\sfd\btheta)
\end{equation}
that can be used to approximate any integrals w.r.t. the true probability measure $\mu_{n}(\sfd\btheta) = p (\btheta | \by_{1:n}) \sfd\btheta$. For example, one can estimate any posterior expectations of the parameter vector $\btheta$ given the observations $\by_{1:n}$, namely
\begin{equation}
\mbE[\btheta | \by_{1:n}] = \int \btheta p(\btheta | \by_{1:n}) \sfd\btheta \simeq \int \btheta \mu_{n}^N (\sfd\btheta) = \frac{1}{N} \sum_{i} \btheta_n^i =: \btheta_{n}^N
\end{equation}
Since we have constructed a complete distribution, statistical errors can be estimated as well. 
The \textit{a posteriori} covariance matrix of vector $\btheta$ can be approximated as 
\begin{equation} \label{eqVarianceTheta}
\mbE\left[
	\left(
		\btheta - \mbE[\btheta|\by_{1:n}]
	\right) \left(
		\btheta - \mbE[\btheta|\by_{1:n}]
	\right)^\top | \by_{1:n}
\right] \simeq \frac{1}{N}\sum_{i=1}^{N} \left( \btheta_n^i - \btheta_n^N \right) \left( \btheta_n^i - \btheta_n^N \right)^\top =: \bP_n^N.
\end{equation}

As a byproduct, Algorithm \ref{alNHF} also yields an approximate predictive pdf for $\bx_n$, namely
\begin{equation}
\nonumber \hat{p}(\bx_{n}| \by_{1:n-1}) = \sum_{i=1}^{N} w_n^i \hat{p}(\bx_{n}| \btheta_n^i, \by_{1:n-1}).
\end{equation}
If one computes the approximate filter, $\hat{p}(\bx_{n}| \btheta_n^i, \by_{1:n-1})$ as well, then the joint probability distribution of $\btheta$ and $\bx_{n}$ conditioned on $\by_{1:n}$ (denoted $\pi_n(\sfd \btheta \times \sfd \bx_{n})$) can be approximated as
\begin{equation}
\nonumber \pi_n^N (\sfd \btheta \times \sfd \bx_{n}) = \sum_{i=1}^{N} w_n^i \hat{p}(\bx_{n}|\btheta, \by_{1:n}) \delta_{\bar \btheta_n^{i}} (\sfd \btheta) \sfd \bx_{n}.
\end{equation}

The scheme of Algorithm \ref{alNHF} is referred to as \textit{nested} because the SMC algorithm generates, at each time step $n$, a set of samples $\{ \btheta_n^1, \ldots, \btheta_n^N \}$ and, for each sample $\btheta_n^i$, we embed a filter in the state space $\Real^{d_x}$ in order to compute the pdf $\hat p(\bx_n | \bar{\btheta}_n^i , \by_{1:n-1} )$ and the approximate likelihood $\hat u_n(\btheta_n^i)$. The term \textit{hybrid} is used because the embedded filters need not be Monte Carlo methods --a variety of techniques can be used and in this paper we focus on Gaussian filters, which are attractive because of their (relative) computational simplicity. A scheme with nested particle filters was thoroughly studied in \cite{Crisan18bernoulli,Crisan17}.

Let us finally remark that the NHF scheme relies on two approximations:
\begin{itemize}
	\item \textit{Jittering of the parameters}: The difficulty of drawing samples from $\mu_{n-1} (\sfd\btheta) = p(\btheta | \by_{1:n-1}) \sfd\btheta$ can be circumvented if we content ourselves with an approximate sampling step. In particular, if we have computed a Monte Carlo approximation $\mu_{n-1}^N (\sfd\btheta) =\frac{1}{N} \sum_{i=1}^{N} \delta_{\btheta_{n-1}^i} (\sfd\btheta)$ at time $n-1$ (with some of the samples replicated because of the resampling step) then we can generate new particles $\bar \btheta_n^i$, $i=1,\ldots,N \sim \kappa_N (\sfd\btheta | \btheta_{n-1}^i)$,
	where $\kappa_N (\sfd\btheta | \btheta')$ is a Markov kernel, i.e., a probability distribution for $\btheta$ conditional on $\btheta'$. See Section \ref{sConvergence} for guidelines on the selection of this kernel. Intuitively, we can either jitter \textit{a few} particles with arbitrary variance (while leaving most of them unperturbed) or jitter \textit{all} particles with a controlled variance that decreases as $N$ increases.

	\item \textit{Estimation of likelihoods}: The sequential approximation of Eqs. \eqref{eqRecur1}--\eqref{eqRecur3} yields biased estimates of the likelihoods $u_n(\btheta_n)$ \cite{Crisan18bernoulli}. This is discussed in Section \ref{sConvergence}. In Appendix \ref{apEnKF} we provide details on the computation of the estimates $\hat p(\bx_n | \bar{\btheta}_n^i , \by_{1:n-1} )$ and $\hat u_n (\btheta_n)$ using an ensemble Kalman Filter (EnKF). Other techniques (e.g., particle filters as in \cite{Crisan18bernoulli} or sigma-point Kalman filters \cite{Ambadan09,Arasaratnam09}) can be used as well.
	
\end{itemize}

%
\subsection{Sequential quasi Monte Carlo hybrid filter} \label{ssSQMC}

The SMC method in the first layer of Algorithm \ref{alNHF} can be replaced by other schemes that rely on the point-mass representation of the posterior probability distribution $\mu_n(\sfd\btheta)$. It is possible to devise procedures based, for instance, on an unscented Kalman filter \cite{Julier04} or other sigma-point Kalman methods \cite{Ambadan09,Arasaratnam09} to obtain a Gaussian approximation of $\mu_n(\btheta)$. Such Gaussian approximations, however, can be misleading when the posterior distribution is multimodal.

In this subsection, we describe a NHF method (hence, of the same class as Algorithm \ref{alNHF}) where the SMC scheme is replaced by a sequential quasi-Monte Carlo (SQMC) procedure of the type introduced in \cite{Gerber15}. The term quasi-Monte Carlo (QMC) refers to a class of deterministic methods for numerical integration \cite{Niederreiter92} that employ low-discrepancy point sets (e.g., Halton sequences \cite{Halton64} or Sobol sequences \cite{Bratley88}), instead of random sample sets, for the approximation of multidimensional integrals. In the context of QMC, \textit{discrepancy} is defined to quantify how uniformly the points in a sequence are distributed into an arbitrary set $S$. Hence, the lowest discrepancy is attained when these points are equi-distributed. The main advantage of (deterministic) QMC methods over (random) Monte Carlo schemes is that they can attain a faster rate of convergence relative to the number of points in the grid, $N$. The main disadvantage is that the generation of low-discrepancy points in high-dimensional spaces can be computationally very costly compared to the generation of random Monte Carlo samples \cite{Martino18}. Within a NHF, the use of QMC should lead to a better performance/complexity trade-off as long as the parameter dimension, $d_\theta$, is relatively small. This is illustrated numerically for a stochastic two-scale Lorenz 96 model in Section \ref{sNumericalResults}.



The NHF based on the SQMC methodology of \cite{Gerber15} can be obtained from Algorithm \ref{alNHF} if we replace the sampling and resampling steps typical of the SMC schemes by the generation of low-discrepancy point sets. Let $\{ \bv_n^i \}_{i=1}^N$ be a Halton sequence of low-discrepancy (deterministic) uniform samples \cite{Halton64}. These uniform samples can be used to generate low-discrepancy variates from other distributions via a number of methods \footnote{One can use a number of techniques used to produce random samples from a given uniform source. See \cite{Martino18} for a comprehensive description of the field, both for single and multivariate distributions.}. For example, the Box-Muller transformation \cite{BoxMuller} can be used to generate pairs of independent, standard, normally distributed pseudo-random numbers. We explicitly indicate the use of low-discrepancy uniform numbers, $\bv_n^i$, in the generation of samples with general distributions by conditioning on $\bv_n^i$. Hence, drawing the $i$-th sample from the prior parameter pdf, $\btheta_0^i \sim p(\btheta)$, is now replaced by $\btheta_0^i \sim p(\btheta|\bv_0^i)$. In order to propagate the $i$-th sample at time $n-1$, $\btheta_{n-1}^i$, into time $n$, we draw from the kernel $\kappa_N(\btheta_n|\btheta_{n-1}^i, \bv_n^i)$. If sampling is needed in the second layer of filters (in order to compute the estimates $\hat p(\bx_n|\bar \btheta_n^i, \by_{1:n-1})$ and $\hat u_n(\bar \btheta_n^i)$) we use additional Halton sequences in a similar way. 

In order to keep the low-discrepancy property across the resampling step, we additionally introduce the following functions (see \cite{Gerber15} for details).
\begin{itemize}
	\item A discrepancy-preserving bijective map $\psi : \Real^{d_x} \rightarrow [0,1]^{d_{\btheta}}$. Several choices are possible for this function. Following \cite{Gerber15}, here we assume
		\begin{equation}
		 \psi(\bar \btheta_n^i)= \left[ 
		 	1 + \exp\left(
				- \frac{\bar \btheta_n^i - \btheta_n^-}{\btheta_n^+ - \btheta_n^-}
			\right) 
		\right]^{-1},
		\label{eqPsieq}
		\end{equation}  
		where $\btheta_n^-$ and $\btheta_n^+$ are the $d_\theta$-dimensional vectors whose $j$-th components are, respectively,
		$$
		[ \btheta_n^- ]_j = m_{n,j}^N - 2 s_{n,j}^{2N} \quad \mbox{and} \quad
		[ \btheta_n^+ ]_j = m_{n,j}^N + 2 s_{n,j}^{2N},
		$$
		whereas $m_{n,j}^N = \sum_{i=1}^N w_n^i [\bar \btheta_n^i]_j$ and $s_{n,j}^{2N} = \sum_{i=1}^N w_n^i \left( [\bar \btheta_n^i]_j - m_{n,j}^N \right)^2$, $j=1, \ldots, d_\theta$, are component-wise means and variances.
	\item The inverse Hilbert curve, $\sfh : [0,1]^{d_{\btheta}} \longrightarrow [0,1]$, which is a continuous fractal space-filling curve that provides a locality-preserving map between a 1-dimensional and a $d_{\btheta}$-dimensional space \cite{Moon01}. 
\end{itemize}
The SQMC-based NHF is outlined in Algorithm \ref{alSQMC-HF}.

		\begin{algoritmo} \label{alSQMC-HF}
	Sequential quasi Monte Carlo nested hybrid filter (SQMC-NHF).
	
	\textbf{Inputs}: 
	\begin{itemize}
		\item[-] Number of Monte Carlo samples, $N$.
		\item[-] \textit{A priori} pdf's  $p(\btheta)$ and $p(\bx_0)$.
		\item[-] A Markov kernel $\kappa_N(\sfd\btheta |  \btheta')$ which, given $\btheta'$, generates \textit{jittered} parameters $\btheta \in \Real^{d_\theta}$.
	\end{itemize}

	\textbf{Procedure}:
	\quad \begin{enumerate}	
		\item Initialization
		
		\begin{enumerate}
			\item Generate QMC uniform samples $\{\bv_{-1}^i, \bv_0^i \}_{i=1}^N$ in $[0,1)^{d_{\btheta}}$. Draw $\btheta_0^{i} \sim p(\btheta | \bv_{-1}^i )$, $i=1, \ldots,N$.

		\end{enumerate}

		\item Recursive step, $n \ge 1$.	
		\begin{enumerate}
			\item For $i=1,\ldots,N$:
			\begin{enumerate}
				\item If $n=1$, then draw $\bar{\btheta}_1^{i} \sim \kappa_N (\sfd\btheta | \btheta_0^i, \bv_0^{i})$, else draw $\bar \btheta_n^i \sim \kappa_N (\sfd\btheta | \btheta_{n-1}^i, \tilde \bv_{n-1}^{\sfc (i)})$, for $n \ge 2$.
				
				\item Approximate $\hat p(\bx_n | \bar{\btheta}_n^i , \by_{1:n-1} )$.
				
				\item Use this approximation to compute the estimate
				\begin{equation}
				\hat{u}_n (\bar{\btheta}_n^i) = \int p(\by_n | \bar{\btheta}_n^i , \bx_n) \hat p(\bx_n | \bar{\btheta}_n^i , \by_{1:n-1} ) \sfd\bx_n.
				\end{equation}
				and let $w_n^i \propto \hat{u}_n(\bar{\btheta}_n^i)$ be the normalized weight of $\bar{\btheta}_n^i$.
			\end{enumerate}
			\item Generate a QMC point set $\{\bv_n^i\}_{i=1}^N$ in $[0,1)^{d_{\btheta}+1}$; let $\bv_n^i = (v_n^i, \tilde \bv_n^i) \in [0,1) \times [0,1)^{d_{\btheta}}$.
			
			\item Hilbert sort: find a permutation $\sf{b}$ such that 
			$$
			\begin{array}{cl}
			(\sf{h} \circ \psi) (\bar \btheta_{n}^{\sfb (1)}) \le \ldots \le (\sf{h} \circ \psi) (\bar \btheta_{n}^{\sfb (N)}), &\mbox{if $d_{\btheta} \ge 2$}\\
			\bar \btheta_{n}^{\sfb (1)} \le \ldots \le \bar \btheta_{n}^{\sfb (N)}, &\mbox{if $d_{\btheta} = 1$}.\\
			\end{array}
			$$
			
			\item Resampling: find a permutation $\sfc$ such that $v_n^{\sfc (1)} \le \ldots \le v_n^{\sfc (N)}$. For $i=1, \ldots, N$, set
			$
			\btheta_n^i = \bar \btheta_n^j
			$
			if, and only if, 
			$$
			\sum_{k=1}^{j-1} w_n^{\sfb(k)} < v_n^{\sfc(i)} \le \sum_{k=1}^j w_n^{\sfb(k)}, \quad j \in \{ 1, \ldots, N \}.
			$$
						
		\end{enumerate}
		
	\end{enumerate}
	
	\textbf{Outputs}: A set of particles $ \{ \btheta_n^i\}_{i=1}^{N} $ and a probability measure $\mu_n^N (\sfd \btheta)=\frac{1}{N} \sum_{i=1}^{N} \delta_{\btheta_n^i} (\sfd \btheta)$.
\end{algoritmo}

\section{Convergence analysis} \label{sConvergence}

The nested filtering schemes of Section \ref{sNHFing} admit various implementations depending on how we choose to approximate the conditional pdf $p(\bx_n|\by_{1:n-1},\btheta)$ which, in turn, is needed to estimate the likelihood function and compute the importance weights $w_n^i \propto \hat{u}(\bar{\btheta}_n^i) \simeq u_n(\bar \btheta_n^i)$, $i=1, \ldots, N$.

For each choice of approximation method, the estimate $\hat u_n(\btheta)$ may behave differently and yield different convergence properties. Here we assume that $\hat u_n(\btheta)$ is a random variable with finite mean $\bar u_n(\btheta) = \mbE[ \hat u_n(\btheta) ] < \infty$ and finite moments up to some prescribed order $p \ge 1$. Specifically, we make following assumption.
  
\begin{Assumption} \label{asLikelihood}
	Given $\btheta \in \Real^{d_\theta}$, the estimator $\hat{u}_n({\btheta})$ is random and can be written as
	\begin{equation}
	\hat{u}_n({\btheta})  = \bar{u}_n({\btheta}) + {m}_n({\btheta}), 
	\end{equation}
	where ${m}_n(\btheta)$ is a zero-mean r.v. satisfying $\mbE[ {m}_n(\btheta)^p ] \le \sigma^p < \infty$ for some prescribed $p \ge 1$. Furthermore, the mean $\bar u_n(\btheta) = \mbE\left[ \hat u_n(\btheta) \right]$ has the form 
	\begin{equation}
	 	\bar{u}_n({\btheta}) = u_n({\btheta}) + b_n({\btheta}),
	\end{equation}
	 where $b_n(\btheta)$ is a deterministic and absolutely bounded bias function.
\end{Assumption}

In the sequel, we use $D \subseteq \Real^{d_\theta}$ to denote the support set of the parameter vector $\btheta$. Given a real function $a:D\rw\Real$, its absolute supremum is indicated as $\| a \|_\infty := \sup_{\btheta\in D} | a(\btheta) |$. The set of absolutely bounded real functions on $D$ is denoted $B(D)$, i.e., $B(D):=\left\{ (a:D\rw\Real) : \| a \|_\infty < \infty \right\}$. For our analysis we assume that $u_n \in B(D)$ and, since we have also assumed the bias function $b_n$ to be bounded, it follows that $\bar u_n \in B(D)$, i.e., $\| \bar u_n \|_\infty < \infty$. To be precise, we impose the following assumption.

\begin{Assumption} \label{asOnG}
	Given a fixed sequence of observations $y_{1:n}$, the family of functions $\{ \bar u_n(\btheta), \btheta \in D \}$ satisfies the following inequalities for each $n=1, 2, ...$:
	\begin{enumerate}
		\item $\bar u_n \in B(D)$, and
		\item $\bar u_n(\btheta) > 0$ for any $\btheta\in D$.
	\end{enumerate}
\end{Assumption}

Since $\| u_n \|_\infty < \infty$, A.\ref{asOnG}.1 follows from assumption A.\ref{asLikelihood}. Similarly, if $u_n(\btheta)>0$ for all $\btheta\in D$ then A.\ref{asOnG}.2 is a natural assumption (since $\hat u(\btheta)$ is an estimator of a positive magnitude).

We shall prove that, because of the bias $b_n(\btheta)$, the approximation $\mu_n^N$ converges to the perturbed probability measure $\bar \mu_n$ induced by the mean function $\bar u_n$, instead of the true posterior probability measure $\mu_n$ induced by the true likelihood function $u_n$. 

To be specific, the sequence of posterior measures $\mu_n$, $n \ge 1$, can be constructed recursively, starting from a prior $\mu_0$, by means of the projective product operation \cite{Bain08}, denoted $\mu_n = u_n \cdot \mu_{n-1}$. When $u$ is a positive and bounded function and $\mu$ is a probability measure, the new measure $u \cdot \mu$ is defined in terms of its integrals. In particular, if $a \in B(D)$ then 
$$
\int a(\btheta) (u\cdot\mu)(\sfd\btheta) := \frac{
	\int a(\btheta) u(\btheta) \mu(\sfd\btheta)
}{
	\int u(\btheta) \mu(\sfd\btheta)
}.
$$
For conciseness, hereafter we use the shorthand
$$
(a,\mu) := \int a(\btheta)\mu(\sfd\btheta)
$$
for the integral of a function $a(\btheta)$ w.r.t. a measure $\mu(\sfd\btheta)$. With this notation, we can write
\begin{equation}
(a,\mu_n) = (a, u_n \cdot \mu_{n-1}) = \frac{
	(au_n,\mu_{n-1})
}{
	(u_n,\mu_{n-1})
}.
\label{eqDefProjProd}
\end{equation}

If, instead of the true likelihood $u_n$, we use the {\em biased} function $\bar u_n=u_n+b_n$ to update the posterior probability measure associated to the parameter vector $\btheta$ at each time $n$ then we obtain the new sequence of measures
$$
\bar \mu_0 = \mu_0, \quad \bar \mu_n = \bar u_n \cdot \bar \mu_{n-1}, \quad n=1, 2, ...,
$$
where, according to the definition of the projective product,
$$
(a,\bar \mu_n) = \frac{
	(a\bar u_n,\bar \mu_{n-1})
}{
	(\bar u_n,\bar \mu_{n-1})
}
$$
for any integrable function $a(\btheta)$. Note that the two sequences, $\mu_n$ and $\bar \mu_n$, start from the same prior $\mu_0$. Obviously, we recover the original sequence, i.e, $\bar \mu_n \rw \mu_n$, when the bias vanishes, $b_n\rw \boldsymbol{0}$.

In this section we prove that the approximation $\mu_n^N$ generated by a generic nested filter that satisfies A.\ref{asLikelihood} and A.\ref{asOnG} converges to $\bar \mu_n$ in $L_p$, for each $n=1, 2, ...$, under additional regularity assumption on the jittering kernel $\kappa_n$. 
\begin{Assumption} \label{asKernel}
	The kernel $\kappa_N$ used in the jittering step satisfies the inequality 
	\begin{equation}
	\sup_{\btheta' \in D} \int | h(\btheta) - h(\btheta') | \kappa_N(\sfd\btheta | \btheta') \le \frac{
		c_\kappa\| h \|_\infty
	}{
		\sqrt{N}
	}
	\label{eqAsKernel1}
	\end{equation}
	for every $h \in B(D)$ and some constant $c_\kappa < \infty$ independent of $N$.
\end{Assumption}

A simple kernel that satisfies A.\ref{asKernel} is \cite{Crisan18bernoulli}
$$
\kappa_N(d\btheta|\btheta') = (1-\epsilon_N)\delta_{\btheta'}(\sfd\btheta) + \epsilon_N \kappa(\sfd\btheta|\btheta'),
$$
where $0 < \epsilon_N \le \frac{1}{\sqrt{N}}$ and $\kappa(\sfd\btheta|\btheta')$ is an arbitrary Markov kernel with mean $\btheta'$ and finite variance, for example $\kappa(\sfd\btheta|\btheta') = \mathcal{N}(\btheta|\btheta',\tilde \sigma^2 \boldsymbol{I}_{d_{\btheta}})$, where $\tilde \sigma^2 < \infty$ and $\boldsymbol{I}_{d_{\btheta}}$ is the identity matrix. Intuitively, this kind of kernel changes each particle with probability $\epsilon_N$ and leaves it unmodified with probability $1-\epsilon_N$. 

Finally, we can state a general result on the convergence of Algorithm \ref{alNHF}. For a real random variable $x$ and $p \ge 1$, let $\| x \|_p$ denote the $L_p$ norm, i.e. $\| x \|_p := \mbE[|x|^p]^{\frac{1}{p}}$.
\begin{theorem} \label{thConvergence}
Let the sequence of observations $y_{1:n_o}$ be arbitrary but fixed, with $n_o < \infty$, and choose an arbitrary function $h \in B(D)$. If assumptions A.\ref{asLikelihood}, A.\ref{asOnG} and A.\ref{asKernel} hold, then
	\begin{equation}
		\| (h, \mu_{n}^N) - (h, \bar \mu_{n}) \|_p \le \frac{
		c_n \| h \|_{\infty}
		}{
		\sqrt{N}
		}, \quad \mbox{for $n = 0, 1, \ldots, n_o$ and any $p\ge 1$},
		\label{eqBoundResampling}
	\end{equation}
where $\{c_n\}_{0 \le n \le n_o}$ is a sequence of finite constants independent of $N$. 
\end{theorem}

\noindent \textbf{Proof:} See Appendix \ref{apConvergence}. $\QED$

We remark that Theorem \ref{thConvergence} does {\em not} state that the approximate posterior probability measure output by Algorithm \ref{alNHF}, $\mu_n^N$, converges to the true posterior measure $\mu_n$, but to the biased version $\bar \mu_n$. Moreover, the latter depends on the choice of filters used in the second layer of Algorithm \ref{alNHF} (i.e., on the estimator of the likelihood, $\hat u_n$). The value of this theorem is that it guarantees the numerical consistency of the the nested hybrid filter: as we increase the computational effort (by increasing $N$), the random probability measure $\mu_n^N$ converges to a well defined limit (and so do any point estimators that we may derive from it, e.g., the posterior mean estimator $\btheta_n^N$). The connection between this limit measure, $\bar \mu_n$, and the true posterior measure $\mu_n$ is given by assumption A.\ref{asLikelihood} and the projective product operation, namely,
$$
\bar \mu_n = (u_n + b_n) \cdot \bar \mu_{n-1}, \quad \mbox{while} \quad \mu_n = u_n \cdot \mu_{n-1}, 
$$    
with both sequences starting with a common prior measure $\bar \mu_0 = \mu_0$. The practical performance of the proposed schemes (with finite $N$) is explored numerically in the sequel.

\section{ A stochastic Lorenz 96 model} \label{sStochasticLorenz}

In order to assess the proposed methods numerically, we have applied them to a stochastic, discrete-time version of the two-scale Lorenz 96 model \cite{Arnold13,Hakkarainen11,Hirata14}. The latter is a deterministic system of nonlinear differential equations that displays some key features of atmosphere dynamics (including chaotic behavior) in a relatively simple model of arbitrary dimension (the number $d_x$ of dynamic variables can be scaled as needed). The model consists of two sets of dynamic variables, $\bx$ and $\boldsymbol{z}$. The system of stochastic differential equations takes the form
\begin{equation}
\begin{split}
\sfd {\bx} &= \bbf_1 (\bx, \boldsymbol{z}, \boldsymbol{\alpha}) \sfd t + \sigma \sfd \boldsymbol{w}_1 \\
\sfd{\boldsymbol{z}} &= \bbf_2 (\bx, \boldsymbol{z}, \boldsymbol{\alpha}) \sfd t + \bar \sigma \sfd \boldsymbol{w}_2
\end{split}
\label{eqslowfastvariables}
\end{equation}
where $\bx(t)$ and $\boldsymbol{z}(t)$ represent the \textit{slow} and \textit{fast} variables, respectively, $\boldsymbol{w}_1$ and $\boldsymbol{w}_2$ are Wiener processes, $\sigma, \bar \sigma > 0$ are known scale parameters and $\boldsymbol{\alpha}$ is a parameter vector of dimension $d_{\alpha} = 4$. Let us assume there are $d_x$ slow variables, ${x}_j$, $j=0, \ldots, d_x-1$, and $L$ fast variables per slow variable, i.e., $z_l$, $l=0, ..., d_xL-1$, overall. The maps $\bbf_1$ and $\bbf_2$ are $\Real^{d_x} \times \Real^L \times \Real^{d_{\alpha}} \rightarrow \Real^{d_x}$ and $\Real^L \times \Real^{d_x} \times \Real^{d_{\alpha}}  \rightarrow \Real^L$ functions, respectively, that can be written (skipping the time index $t$) as
\begin{equation} 
\begin{split}
\bbf_1 = [f_{1,0}, \ldots, f_{1,d_x-1}]^\top \quad &\text{and} \quad f_{1,j} (\bx, \boldsymbol{{z}},\boldsymbol{\alpha}) = -x_{j-1} ( x_{j-2} - x_{j+1}) - x_{j} + F - \frac{H C}{B} \sum_{l=(j-1)L}^{Lj-1} z_{l},  \\
\bbf_2 = [f_{2,0}, \ldots, f_{2,d_xL-1}]^\top \quad &\text{and} \quad f_{2,l} (\boldsymbol{z},\boldsymbol{\alpha}) = -CB  z_{l+1} (  z_{l+2} -  z_{l-1} ) - C  z_l + \frac{C F}{B} + \frac{HC}{B} x_{\lfloor \frac{l-1}{L} \rfloor}, 
\end{split}
\label{eqF1F2}
\end{equation}
where $\boldsymbol{\alpha}$ contains the parameters $F$,$C$,$H$ and $B$. The forcing parameter $F$ controls the turbulence of the chaotic flow, $C$ determines the time scale of the fast variables $\{z_l\}_{l\ge 0}$, $H$ controls the strength of the coupling between the fast and slow variables and $B$ determines the amplitude of the fast variables \cite{Arnold13}. The dynamic variables are assumed to be arranged on a circular structure, hence the operations on the $j$ indices are modulo $d_x$ and operations on the $l$ indices are modulo $L$. This means that for any integer $k$, $j+k \equiv (j+k) \mbox{ mod } d_x$ and $l+k \equiv (l+k) \mbox{ mod } L$. Notation $\lfloor a \rfloor$ indicates the truncation of a positive real number $a$ to the closest integer smaller than $a$.

We apply the 4th order Runge-Kutta (RK4) method \cite{Gard88} to obtain a discrete-time version of the two-scale Lorenz 96 model. To be specific, we numerically integrate Eq. \eqref{eqslowfastvariables} by means of the stochastic difference equations
\begin{equation} 
\begin{split}
{\bx}_n & = {\bx}_{n-1} + {\bbF}_1^m({\bx}_{n-1},{\boldsymbol{z}}_{n-1}, \boldsymbol{\alpha},h,\sigma \bv_n), \\
{\boldsymbol{z}}_n & = {\boldsymbol{z}}_{n-1} + {\bbF}_2^m({\bx}_{n-1},{\boldsymbol{z}}_{n-1}, \boldsymbol{\alpha},h,\bar \sigma \bar{\bv}_n)
\end{split}
\label{eqRK4f1f2}
\end{equation}
where $h>0$ is the integration step-size, and $\bv_n$ and $\bar{\bv}_n$ are sequences of i.i.d. standard Gaussian r.v.'s. 

We assume that the observations are linear but can only be collected from this system once every $T$ discrete-time steps. Moreover, only 1 out of $K$  slow variables can be observed. Therefore, the observation process has the form
\begin{equation}
\by_n = 
\left[ \begin{array}{c}
x_{K,nT} \\
x_{2K,nT}\\
\vdots \\
x_{d_yK,nT}
\end{array} \right] + \boldsymbol{r}_n,
\label{eqObservations}
\end{equation}
where $n=1, 2, ...$ and $\boldsymbol{r}_n$ is a sequence of  i.i.d. r.v.'s with common pdf $\mN(\boldsymbol{r}_n | 0, \sigma_o^2 \textbf{I}_{d_y})$. 

In our computer experiments, system \eqref{eqF1F2} is often employed to generate both ground-truth values for the slow variables $\{\bx_{n}\}_{n \ge 0}$ and synthetic observations, $\{ \by_n \}_{n \ge 1}$. As a forecast model for the slow variables it is common \cite{Hakkarainen11} to use the SDE
\begin{equation} \label{eqContinuousAnsatz}
\sfd{x}_j = f_j (\bx, \btheta) \sfd t + \sigma \sfd w_j = [-x_{j-1} ( x_{j-2} - x_{j+1}) - x_{j} + F - \ell(x_{j},\sfa)] \sfd t + \sigma \sfd w_j, \quad j=0, ..., d_x-1,
\end{equation}
where $\sfa=[\sfa_1,\sfa_2]^\top$ is a (constant) parameter vector, $\btheta = [ F, \sfa^\top ]^\top$ contains all the parameters, function $\ell(x_{j},\sfa) \in \Real$ is a polynomial ansatz for the coupling term $\frac{HC}{B} \sum_{l=(j-1)L}^{Lj-1} z_{l}$ in \eqref{eqF1F2} and $w_j$ is a standard Wiener process. In this paper we assume that $\ell(x_{j},\sfa)$ is a polynomial in $x_j$ of degree $2$, characterized by the coefficients $\sfa_1$ and $\sfa_2$ as
\begin{equation} 
\ell(x_{j},\sfa) = \sfa_1 x_j^2 + \sfa_2 x_j . \nonumber
\end{equation}
Then, the system \eqref{eqRK4f1f2} can be replaced by
\begin{equation} \label{eqRK4approx}
\boldsymbol{{x}}_n = \boldsymbol{{x}}_{n-1} + \bar{\bbF}^m(\boldsymbol{{x}}_{n-1},\btheta,h,\sigma \bv_n) 
\end{equation}
where $\bar{\bbF}^m$ is the RK4 approximation of the function $\bbf=[f_0,\ldots,f_{d_x-1}]^\top$ in Eq. \eqref{eqContinuousAnsatz}.

 Assuming $\by_n$ is Gaussian distributed and $\boldsymbol{r}_n$ is a sequence of independent and identically distributed noise terms with Gaussian probability distribution, $p(\boldsymbol{r}) = \mathcal{N}( \boldsymbol{r} | 0,\sigma_o^2 \boldsymbol{I}_{d_y})$, then
\begin{equation} 
p(\by_n | \bx_n, \btheta) = \mathcal{N}(\by_n | \bx_{n},\sigma_o^2 \boldsymbol{I}_{d_y})
\end{equation}
which denotes a $d_y$-dimensional Gaussian density with zero mean and covariance matrix $\sigma_o^2 \boldsymbol{I}_{d_y}$, where $\boldsymbol{I}_{d_y}$ is the $d_y \times d_y$ identity matrix.


\section{Numerical results} \label{sNumericalResults}

We have conducted computer simulations to illustrate the performance of the proposed NHF methods. In particular, we have carried out computer experiments for six different schemes: the NPF of \cite{Crisan18bernoulli}, the two-stage filter of \cite{Santitissadeekorn15} and four NHFs that rely on the SQMC and the SMC, both in combination with EKFs or EnKFs. Then, two different versions of Algorithm \ref{alNHF} (SMC-EKF, SMC-EnKF) and  Algorithm \ref{alSQMC-HF} (SQMC-EKF, SQMC-EnKF) are simulated. The simulation setup is described below, followed by the discussion of our numerical results in Section \ref{sResults}.

\subsection{Simulation setup} \label{sSimulationSetup}

For our computer experiments we have used the two-scale Lorenz 96 model of Eq. \eqref{eqslowfastvariables}, in order to generate 
\begin{itemize} 
	\item reference signals $\tilde{\bx}_k$, $k=0,1,\ldots$, used as ground truth for the assessment of the estimators, and 
	
	\item sequences of observations, $\by_n$, $n=1,2,\ldots$ as in Eq. \ref{eqPsieq}.
\end{itemize}

The model is integrated using the RK4 method with Gaussian perturbations \cite{Gard88} (as outlined in Eq. \eqref{eqRK4f1f2}). The integration step is set to $h=5 \times 10^{-3}$ continuous-time units through all experiments and the fixed model parameters are $F = 8$, $H = 0.75$, $C = 10$ and $B = 15$. For all experiments, we assume that there are $L = 10$ fast variables per slow variable, hence the total dimension of the model is $10d_x$ (with different values of $d_x$ for different experiments). The noise scaling factors are $\sigma = \frac{h}{4} = 0.25 \times 10^{-3}$ and $\sigma_o = 4$, both assumed known. We assume that half of the slow variables are observed in Gaussian noise, i.e., $K = 2$. 

We assess the accuracy of the estimation algorithms in terms of the mean square error (MSE) of the predictors of the dynamic variables. For the NHFs, these estimators take the form 
\begin{equation}
	\hat{{\bx}}_n = \sum_{i=1}^{N} w_n^i \hat{\bx}_n^i,
\end{equation} 
where $\hat{\bx}_n^i$ is the posterior-mean estimate obtained from the approximate filter $\hat p(\bx_n | \by_{1:n}, \bar \btheta_n^i)$, that can be expressed as $\mathcal{N}(\bx_n | \hat \bx_n^i, \bP_n^i)$, since the approximation is Gaussian. In the plots, however, we show the empirical MSE per dimension resulting directly from the simulations,
\begin{equation}
	\mbox{MSE}_n = \frac{1}{d_x} \parallel {\bx}_n - \hat{\bx}_n \parallel^2.
\end{equation}
averaged over 100 independent simulation runs, being all of them of 40 continuous-time units of duration.

The simulations presented below include running times for the different methods. They have been coded in Matlab R2016a and run on a computer with 64 GB of DRAM and equipped with two Intel Xeon E5-2680 processors (running at 2.80GHz) with 10 cores each and HyperThreading as well as an Intel Xeon Phi co-processor.

\subsection{Results} \label{sResults}

Table \ref{table:NPFEKFEnKF} shows a comparison of the performance of the NPF, the two-stage filter and the four NHFs, based on the use of SMC, SQMC, EKF and EnKF schemes as described in Section \ref{sNHFing}, in terms of their running times and the MSE of the state estimators (averaged over time and dimensions). We have carried out this computer simulation for a model with dimension $d_x = 40$ and a gap between observations of $hm = 0.05$ continuous-time units. All NHFs algorithms work with $N=100$ particles for the approximation of the posterior distributions of the fixed parameters, using $M = d_x = 40$ samples per each EnKF in the second-layer. It can be seen that the highest error is achieved by the NPF, followed by the two-stage filter method. The NPF is also the algorithm that takes the longest running time. Both NHFs using EKF attain the least MSE with the smallest running time. In order to improve the performance of the NPF, the numbers of particles $M$ and $N$ would have to be considerably increased, but this would increase the running times correspondingly (the complexity of the NPF is $\mathcal{O}(NM)$ \cite{Crisan18bernoulli}). 

\begin{table}[htb]	
	\begin{center}
		\begin{tabular}{ | l | c | c |}
			\hline
			Algorithm & Running time (minutes) & MSE \\ \hline
			NHF: SQMC-EKF & 2.16 & \quad 0.46 \quad \\ \hline 
			NHF: SMC-EKF & 2.27 & \quad 0.49 \quad \\ \hline 
			NHF: SQMC-EnKF & 6.83 & \quad 0.62 \quad \\ \hline 
			NHF: SMC-EnKF & 7.12 & \quad 0.95 \quad \\ \hline 
			Two-Stage Filter $(N = 600, M = 400)$ & 6.85 & \quad 4.59 \quad \\ \hline 
			NPF $(N = M = 800)$ & 17.96 & \quad 11.91 \quad  \\ %
			\hline
		\end{tabular}
	\end{center}
	\caption{Running times and average MSE (over time and state dimensions) for the NPF, the two-stage filter and four NHFs, based on the SQMC, the SMC, the EKF and the EnKF, respectively.}
	\label{table:NPFEKFEnKF}
\end{table}

\begin{figure}[htb]
	\begin{subfigure}{0.80\linewidth}
		\includegraphics[width=0.65\linewidth]{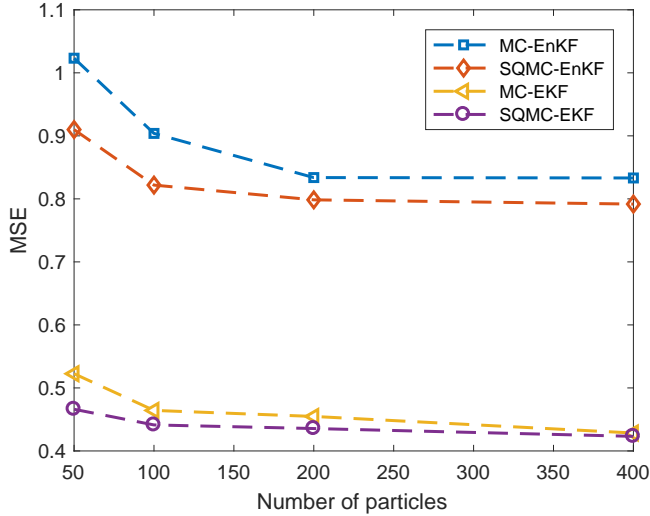}		
\end{subfigure}
\caption{MSE of the different NHFs depending on the number of particles used in the first-layer of the filter. }
\label{msealgorithmsvsparticles}
\end{figure}

In the next experiment we assess the performance of the different NHFs depending on the number of particles used in the first-layer of the filter, in order to choose appropriately this number to carry out the following computer experiments. For this purpose, we consider a model with dimension $d_x = 100$, a gap between observations of $hm = 0.05$ continuous-time units and a number of particles that ranges from $50$ to $400$. Figure \ref{msealgorithmsvsparticles} shows the numerical results for this experiment. We observe that the MSE for the four algorithms stabilizes quickly. At the sight of these results, we set $N=100$ for all remaining experiments. Additionally, Figure \ref{msealgorithmsvsparticles} also shows the difference between the NHFs. Specifically, we see that using SQMC in the first-layer we can slightly improve slightly the performance. For this reason, in the next experiments we only simulate NHFs that rely on SQMC. Moreover, it is easy to observe that the filters that use EKFs in the second-layer obtain better results.

In the next set of computer experiments we compare the SQMC-EKF and the SQMC-EnKF methods in terms of their average MSE and their running times for different values of the state dimension $d_x$ and the gap between consecutive observations $m$ (in discrete time steps). For each combination of $d_x$ and $m$ we have carried out 100 independent simulation runs. The number of particles in the parameter space is fixed, $N=100$, for all simulations, but the size of the ensemble in the EnKFs is adjusted to the dimension, in particular, we set $M=d_x$. 

\begin{figure}[htb]
	\begin{subfigure}{0.49\linewidth}
		\includegraphics[width=0.65\linewidth]{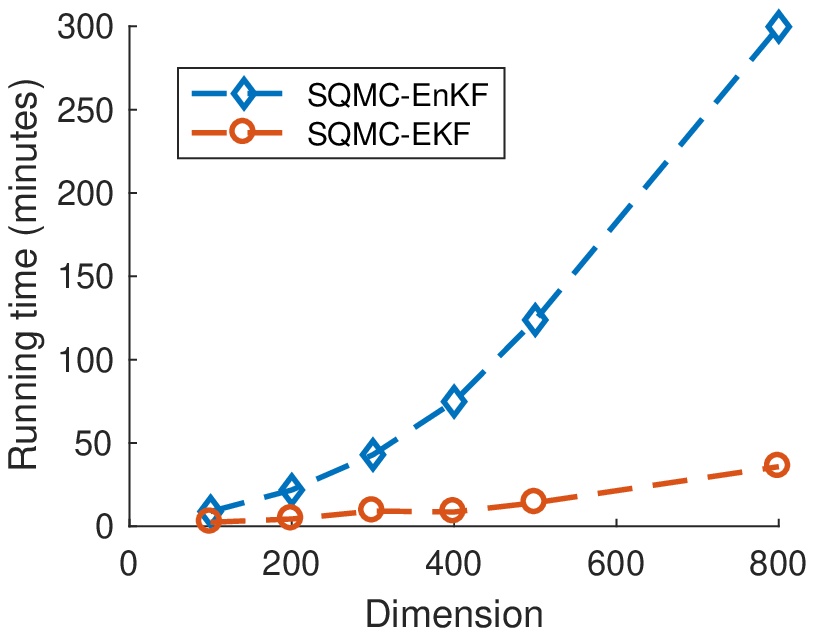}
		\caption{}
		\label{fDIMvsTIME_nosc}
	\end{subfigure}
	\begin{subfigure}[htb]{0.49\linewidth}
		\includegraphics[width=0.65\linewidth]{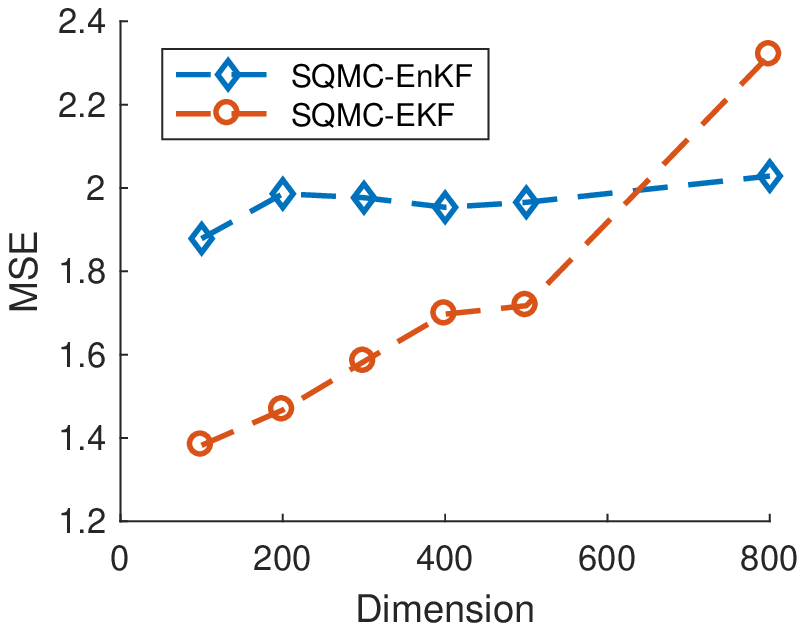}
		\caption{}
		\label{fMSEvsDIM_nosc}
	\end{subfigure}
	\caption{Comparison of the SQMC-EKF (red lines) and SQMC-EnKF (blue lines) in terms of their running time (a) and their $MSE$ (b) as the state dimension $d_x$ increases, with a fixed gap between observations of $T=20$ discrete time steps.}
	\label{fChangingNOSC}
\end{figure}

Figure \ref{fChangingNOSC} shows (a) the running times and (b) the average MSE attained by the two SQMC NHFs when the state dimension $d_x$ ranges from $100$ to $800$. The gap between observations is fixed to $m=20$ (i.e., $0.1$ time units versus $0.05$ in Figure \ref{msealgorithmsvsparticles}). We observe that the SQMC-EKF method attains significantly lower running times compared to the SQMC-EnKF, since the former increases linearly with dimension while the latter increases its cost exponentially. However, the SQMC-EKF obtains an MSE which increase with the dimension $d_x$, while the values of MSE for the SQMC-EnKF method are steady w.r.t. $d_x$. 

\begin{figure}[htb]
	\begin{subfigure}[htb]{0.49\linewidth}
		\includegraphics[width=0.65\linewidth]{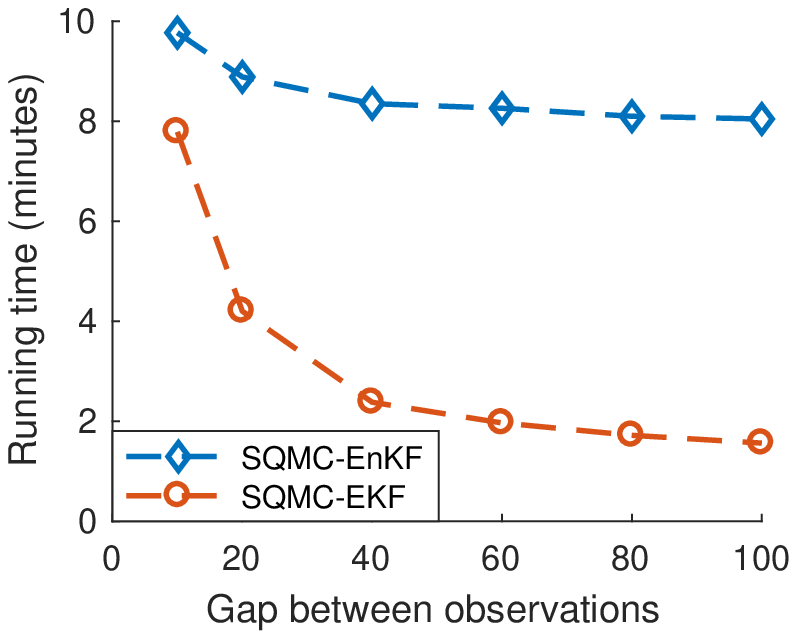}
		\caption{}
		\label{fTOBSvsTIME_tobs}
	\end{subfigure}
	\begin{subfigure}{0.49\linewidth}
		\includegraphics[width=0.65\linewidth]{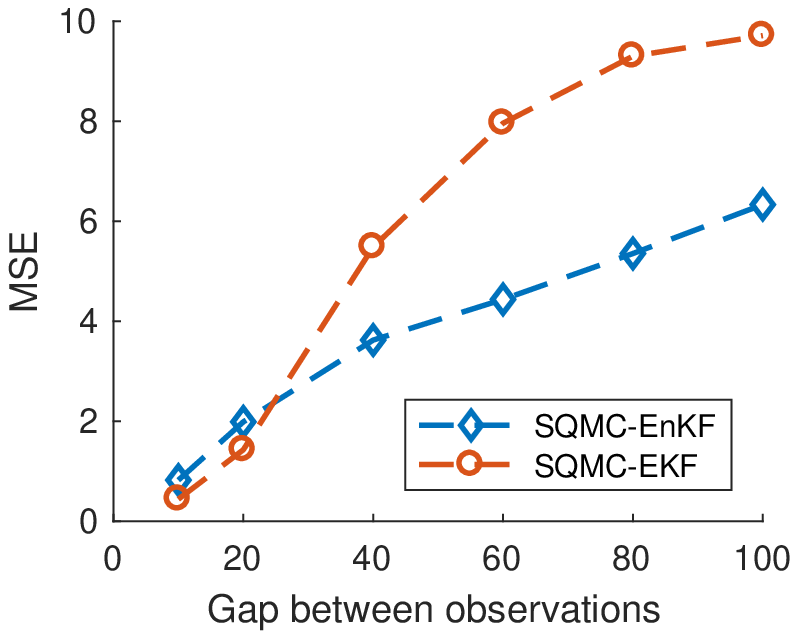}
		\caption{}
		\label{fMSEvsTOBS_tobs}
	\end{subfigure}
	\caption{Comparison of the SQMC-EKF (red lines) and SQMC-EnKF (blue lines) in terms of their running time (a) and their $MSE$ (b) as the gap between observations $m$ increases, with fixed state dimension $d_x=100$. }
	\label{fChangingTOBS}
\end{figure}

Next, Figure \ref{fChangingTOBS} displays the running times and the average MSEs attained by the two NHFs as we increase the gap between observations from $m=10$ to $m=100$ (hence, from $hm=0.05$ to $hm=0.50$ continuous time units). The dimension of the state for this experiment is fixed to $d_x=100$. Note that, as the gap $m$ increases, less data points are effectively available for the estimation of both the parameters and the states. We observe, again, that the SQMC-EnKF is computationally more costly than the SQMC-EKF, however it attains a consistently smaller MSE when the gap between observations increases, suggesting that it may be a more efficient algorithm in data-poor scenarios.  



Finally, we show results for a computer experiment in which we have used the SQMC-EnKF method to estimate the parameters $F$ and ${\sf a}$ and track the state variables of the two-scale Lorenz system with dimension $d_x = 5,000$ and a gap between consecutive observations of $hm=0.05$ continuous-time units. As in the rest of computer simulations, the number of particles used to approximate the sequence of parameter posterior distributions is $N=100$. 

\begin{figure}[htb]
	\begin{subfigure}{0.49\linewidth}
		\includegraphics[width=0.65\linewidth]{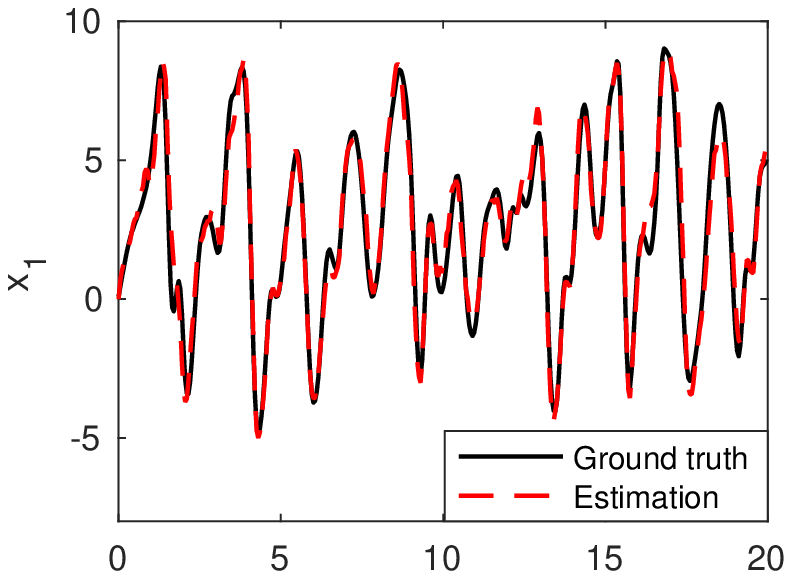}
		\caption{}
		\label{fx1}
	\end{subfigure}
	\begin{subfigure}[htb]{0.49\linewidth}
		\includegraphics[width=0.65\linewidth]{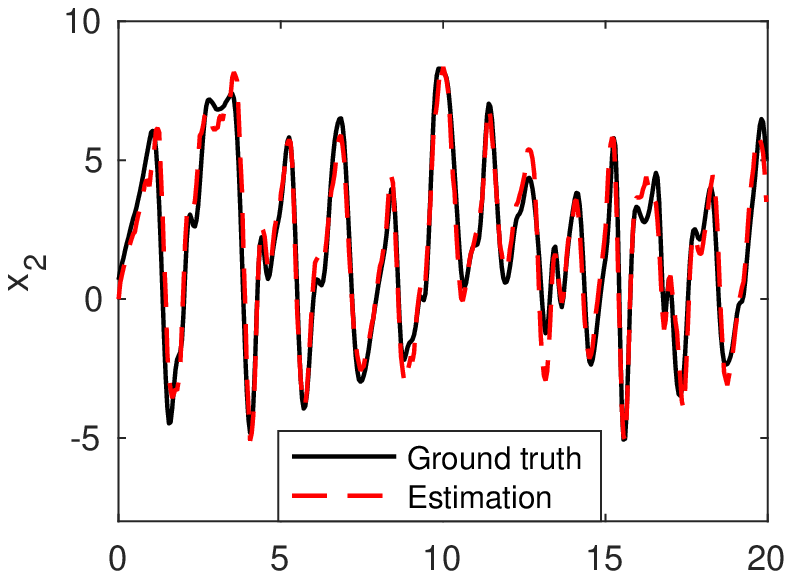}
		\caption{}
		\label{fx2}
	\end{subfigure}
	\caption{Sequences of state values (black line) and estimates (dashed red line) in $x_1$ (a) and $x_2$ (b) over time. Variable $x_1$ is observed (in Gaussian noise), while $x_2$ is unobserved.}
	\label{fX1X25000osc}
\end{figure}

Figure \ref{fX1X25000osc} shows the true state trajectories, together with their estimates, for the first two slow state variables of the two-scale Lorenz 96 model. We note that the first variable, $x_1(t)$, is observed in Gaussian noise (with $\sigma_o=4$) while the second variable, $x_2(t)$, is not observed. The accuracy of the estimation is similar, though, over the 20 continuous-time units of the simulation run (corresponding to $20\times 10^3$ discrete time steps), achieving and $\mbox{MSE} \approxeq 0.87$. Taking into account the steadiness of MSE w.r.t. dimension of SQMC-EnKF in Figure \ref{fMSEvsDIM_nosc} and the values of MSE shown in Figure \ref{fMSEvsTOBS_tobs} for the gap selected in this experiment ($m = 10$), the results obtained are within the expected range.

\begin{figure}[htb]
	\begin{subfigure}{0.49\linewidth}
		\includegraphics[width=0.65\linewidth]{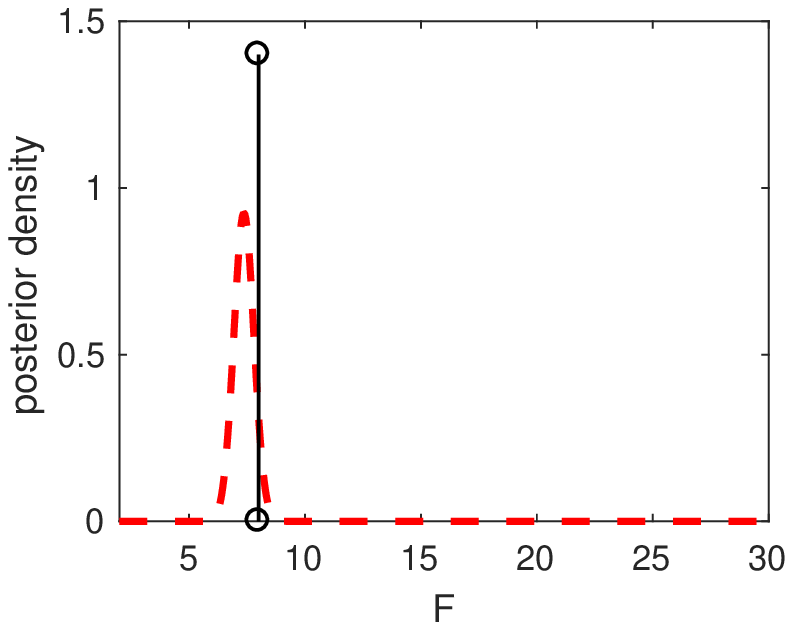}
		\caption{}
		\label{fF}
	\end{subfigure}
	\begin{subfigure}[htb]{0.49\linewidth}
		\includegraphics[width=0.65\linewidth]{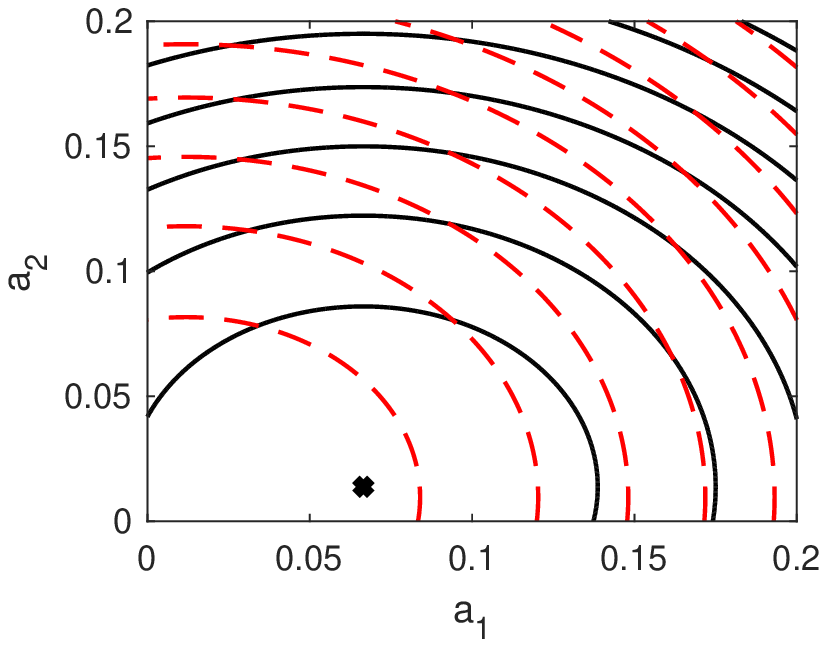}
		\caption{}
		\label{fA}
	\end{subfigure}
	\caption{Posterior density of the parameters ${\sf a}=[{\sf a}_1,{\sf a}_2]^\top$ and $F$ at $t = 5$ in a 5,000-dimensional Lorenz 96 model (red dashed lines). The reference values are represented in black lines.} 
	\label{fCoefficients4000nosc}
\end{figure}

In Figure \ref{fCoefficients4000nosc} we observe the estimated posterior pdf's of the fixed parameters $F$, ${\sf a}_1$ and ${\sf a}_2$, together with the reference values. Note that the value $F=8$ is ground truth, but the values of ${\sf a}_1$ and ${\sf a}_2$ are genie-aided least-squares estimates obtained by observing directly the fast variables of the two-scale model. Figure \ref{fF} displays the true value $F=8$ (vertical line) together with the approximate posterior pdf generated by the same Euler algorithm. We observe that nearly all probability mass is allocated close to the true value. In Fig. \ref{fA} we compare the approximate pdf of the coefficients $ \sf {a} = [\sf a_1, \sf a_2]^T$ produced by the NHF with a kernel density estimator computed from the least-squares genie-aided estimates computed over 100 independent simulations with the same setting. The modes of the two pdf's are slightly shifted but the two functions are otherwise similar. The genie-aided estimate of $\sf {a}$ is located in a light probability region of the density function computed by the NHF.

\section{Conclusions} \label{sConclusions}

We have introduced a nested filtering methodology to recursively estimate the static parameters and the dynamic variables of nonlinear dynamical systems. The proposed framework combines a recursive Monte Carlo approximation method to compute the posterior probability distribution of the static parameters with a variety of filtering techniques to estimate the posterior distribution of the state variables of the system. In particular, we have investigated the use of Gaussian filters, as they admit fast implementations that can be well suited to high dimensional systems. As a result, we have proposed a class of nested hybrid filters that combine Monte Carlo and quasi Monte Carlo schemes for the (moderate dimensional) unknown static parameters of the dynamical system with either extended Kalman filtering or ensemble Kalman filtering for the (higher dimensional) time-varying states. Additionally, when sequential MC is applied in the first layer of the NHF scheme, we have proved that the algorithm converges with rate $\mO\left(N^{-\frac{1}{2}}\right)$ to a well defined limit distribution. We have presented numerical results for a two-scale stochastic Lorenz 96 system, a model commonly used for the assessment of data assimilation methods in the Geophysics. We illustrate the average performance of the methods in terms of estimation errors and running times, and show numerical results for a 5,000-dimensional system. This has been achieved with a relatively inefficient implementation of the method running on a desktop computer, hence we expect that the method can be applied to much larger scale systems using adequate hardware and software.

\section*{Acknowledgments}
This research has been partially supported by the Spanish Ministry of Economy and Competitiveness (projects TEC2015-69868-C2-1-R ADVENTURE and TEC2017-86921-C2-1-R CAIMAN) and the Office of Naval Research (ONR) Global (Grant Award no. N62909-15-1-2011). 


\appendix

\section{Nested hybrid filter implementation using a bank of ensemble Kalman filters} \label{apEnKF}

In this appendix we outline a version of the NHF that employs the ensemble Kalman filter (EnKF) \cite{Evensen03} in order to compute the posterior (approximate) pdf's $\hat p(\bx_n|\by_{1:n-1},\bar \btheta_n^i)$ and $\hat p(\bx_n|\by_{1:n},\bar \btheta_n^i)$, which are needed to evaluate the importance weights $w_n^i \propto \hat u_n(\bar \btheta_n^i)$. In the EnKF, the approximate filter $\hat p(\bx_n | \by_{1:n}, \bar \btheta_n^i)$ is represented by an ensemble of $M$ Monte Carlo particles $\{ \bx_n^{i,j} \}_{j=1}^{M}$, which can be combined to yield an empirical covariance matrix $\bP_n^i$. 

Each ensemble can be stored in a $d_x \times M$ matrix $\boldsymbol{X}_n^i = [ \bx_n^{i,1}, \bx_n^{i,2}, \ldots, \bx_n^{i,M} ]$. The $i$-th mean and the $i$-th covariance matrix can be computed as
$$
\bar{\bx}_n^i = \frac{1}{M} \boldsymbol{X}_n^i \boldsymbol{\mathrm{1}}
\quad \mbox{and} \quad
\bar{\boldsymbol{P}}_n^i = \frac{1}{M} \tilde{\boldsymbol{X}}_n^i (\tilde{\boldsymbol{X}}_n^i)^\top,
$$
respectively, where $\boldsymbol{\mathrm{1}} = [1, \ldots, 1]^\top$ is an $M$-dimensional column vector and $\tilde{\boldsymbol{X}}_n^i = \boldsymbol{{X}}_n^i -  \bar{\bx}_n^i \boldsymbol{\mathrm{1}}^\top$ is an ensemble of deviations from $\bar{\bx}_n^i$. We hence write $\mathcal{N} (\bx_n | \boldsymbol{X}_n^i)$ as a shorthand for the pdf $\mathcal{N} (\bx_n | \bar{\boldsymbol{{x}}}_n^i, \bar{\boldsymbol{{P}}}_n^i)$. 

We assume that the prior pdf of the state is Gaussian with known mean and covariance matrix, namely
\begin{equation}
	p(\bx_0) = \mathcal{N} (\bx_0 | \bar{\boldsymbol{{x}}}_0, \bar{\boldsymbol{{P}}}_0)
\end{equation}
The noise terms in the state space model are also assumed Gaussian, with zero mean and known covariance matrices,
\begin{equation}
\bv_k \sim \mathcal{N}(\bv_k | \boldsymbol{0}, \boldsymbol{Q}) \quad \text{and} \quad \boldsymbol{r}_k \sim \mathcal{N}(\boldsymbol{r}_k | \boldsymbol{0}, \boldsymbol{T}).
\end{equation}

The NHF constructed around a bank of EnKFs is outlined in Algorithm \ref{alEnKF} below.

\begin{algoritmo} \label{alEnKF}
	NHF via EnKF.
	
	\begin{enumerate}
		\item \label{enkf1} Initialization: draw N i.i.d. particles $\btheta_0^{i} \sim \mu_0(\sfd\btheta)$ and $\{\bar{\bx}_0^{i,j}\} \sim p(\bx_0), i=1,\ldots,N$, $j=1,\ldots, M$. Let $\boldsymbol{{X}}_0^i = [ \bx_0^{i,1}, \ldots, \bx_0^{i,M} ]$, $i=1,\ldots,N$.
		
		\item \label{enkf2} Recursive step: at time $n-1$, we have obtained $\mu_{n-1}^N (\sfd\btheta) = \frac{1}{N} \sum_{i=1}^{N} \delta_{\btheta_{n-1}^i} (\sfd\btheta)$ and, for each $i=1,\ldots,N$, $\hat p(\bx_{n-1} | \by_{1:n-1}, \btheta_{n-1}^i) = \mathcal{N} (\bx_{n-1} | \bar{\boldsymbol{{X}}}_{n-1}^i)$.
		
		\begin{enumerate}
			\item  \label{enkf2a} Prediction step: \begin{enumerate}
				\item \label{enkf2ai} Draw $\bar{\btheta}_n^i \sim \kappa_N(\sfd\btheta | \btheta_{n-1}^i)$, $i=1,\ldots,N$. 
				
				\item \label{enkf2aii} For each $i=1,\ldots,N$ compute
				\begin{equation}
				\hat{\boldsymbol{X}}_n^i = {\boldsymbol{{F}}}^m({\boldsymbol{X}}_{n-1}^i , \bar{\btheta}_n^i,h,\sigma \boldsymbol{V}_n^i)
				\end{equation}
				where $\boldsymbol{{V}}_n^i = [ \bv_n^{i,1}, \ldots, \bv_n^{i,M} ]$, $i=1,\ldots,N$, is a $m q d_x \times M$ matrix of Gaussian perturbations ($q$ denotes the order of the underlying RK integrator).
				
				\item \label{enkf2aiii} Set $\hat p(\bx_n|\by_{1:n-1},\bar \btheta_n^i) = \mathcal{N} (\bx_n | \hat{\boldsymbol{X}}_n^i)$.
				
			\end{enumerate}
			
			\item \label{enkf2b} Update step: \begin{enumerate}
				\item \label{enkf2bi} For $i=1,\ldots,N$, compute
				\begin{align}
				\bar{\boldsymbol{M}}_n^i & = \frac{1}{M} \tilde{\boldsymbol{X}}_n^i (\tilde{\boldsymbol{Z}}_n^i)^\top \\
				\bar{\boldsymbol{S}}_n^i & = \frac{1}{M} \tilde{\boldsymbol{Z}}_n^i (\tilde{\boldsymbol{Z}}_n^i)^\top + \boldsymbol{T} \\
				\bar{\boldsymbol{K}}_n^i & =  \bar{\boldsymbol{M}}_n^i (\bar{\boldsymbol{S}}_n^i)^{-1} \label{eqKgainEnKF} \\
				\check{\boldsymbol{X}}_n^i & = \hat{\boldsymbol{X}}_{n}^i + \bar{\boldsymbol{K}}_n^i (\by_n \boldsymbol{\mathrm{1}}^\top - \bar{\boldsymbol{Y}}_n^i)
				\end{align}
				where $\boldsymbol{T} = \sigma_o^2 \boldsymbol{I}_{d_y}$ is the measurement noise covariance, $\bar{\by}_n = \frac{1}{M} \bar{\boldsymbol{Y}}_n^i \boldsymbol{\mathrm{1}}$ and $\bar{\bx}_n^i = \frac{1}{M} \hat{\boldsymbol{X}}_n^i \boldsymbol{\mathrm{1}}$, with $\bar{\boldsymbol{Y}}_n^i = \bg(\hat{\boldsymbol{X}}_n^i, \btheta) + \boldsymbol{R}_n^i$ and $\boldsymbol{R}_n^i = [\boldsymbol{r}_n^{1},  \ldots,  \boldsymbol{r}_n^{M} ]$ a matrix of Gaussian perturbations. $\tilde{\boldsymbol{{X}}}_n^i$ and $\tilde{\boldsymbol{Z}}_n^i$ are calculated as 
				\begin{align}
				\tilde{\boldsymbol{X}}_n^i & = \hat{\boldsymbol{X}}_n^i - \bar{\bx}_n^i \boldsymbol{\mathrm{1}}^\top \\
				\tilde{\boldsymbol{Z}}_n^i & = \frac{1}{M} \bg(\hat{\boldsymbol{X}}_n^i, \btheta) - \bar{\by}_n^i \boldsymbol{\mathrm{1}}^\top
				\end{align}			
				
				\item  \label{enkf2bii} Compute $\hat{u}(\bar{\btheta}_n^i) = \mathcal{N} (\by_n | \bg(\bar{\bx}_n^i, \bar{\btheta}_n^i) , \bar{\boldsymbol{S}}_n^i)$ and obtain the normalized weights, 
				\begin{equation}
				w_i = \frac{\hat{u}(\bar{\btheta}_n^i)}{\sum_{j=1}^{N} \hat{u}_n (\bar{\btheta}_n^j)}, \quad i=1,\ldots,N.
				\end{equation}
				
				\item \label{enkf2biii} Set the filter approximation 
				\begin{equation}
				\hat p(\bx_n | \by_{1:n}, \bar{\btheta}_n^i) = \mathcal{N} (\bx_n | \check{\boldsymbol{X}}_n^i).
				\end{equation}
				
			\end{enumerate}
			\item \label{enkf2c} Resampling: draw indices $j_1, \ldots, j_N$ from the multinomial distribution with probabilities $w_n^1, \ldots, w_n^N$, then set
			\begin{equation}
			\btheta_n^i = \bar{\btheta}_n^{j_i}, \quad \text{and} \quad {\boldsymbol{{X}}}_n^i = \check{\boldsymbol{X}}_n^{j_i}
			\end{equation}
			for $i=1,\ldots, N$. Hence 
			$$
			\hat p(\bx_n|\by_{1:n}, \btheta_n^i) = \mathcal{N} (\bx_n | {\boldsymbol{{X}}}_n^i) \quad 
			\mbox{and} \quad
			\mu_n^N(\sfd\btheta) = \frac{1}{N}\sum_{i=1}^N \delta_{\btheta_n^i}(\sfd\btheta).
			$$
			
		\end{enumerate}
		
	\end{enumerate}

\end{algoritmo}

A computationally expensive step is the inversion of the observation covariance matrix $\bar{\boldsymbol{S}}_n^i$ in step \ref{enkf2bi} and we use the approximation described in Appendix \ref{sInverseSimpl} to alleviate the cost. We note that Algorithm \ref{alEnKF} does not require the computation of the predictive state covariance matrices.


\section{Proof of Theorem \ref{thConvergence}} \label{apConvergence}

%
\subsection{Outline of the proof}

We need to prove that the approximation $\mu_n^N$ generated by a generic nested filter that satisfies assumptions A.\ref{asLikelihood}, A.\ref{asOnG} and A.\ref{asKernel} converges to $\bar \mu_n$ in $L_p$, for each $n=1, 2, ..., n_0 < \infty$. We split the analysis of the nested filter in three steps: jittering, weight computation and resampling. The approximation $\mu_{n-1}^N$ of $\bar \mu_{n-1}$ is available at the beginning of the $n$-th time step. After jittering, we obtain a new approximation,
\begin{equation}
	\check{\mu}_{n-1}^N = \frac{1}{N} \sum_{i=1}^{N} \delta_{\bar{\btheta}_n^i},
\end{equation}
that can be proved to converge to $\bar \mu_{n-1}$ using an auxiliary result from \cite{Crisan18bernoulli}. After the computation of the weights, the measure
\begin{equation}
	\tilde{\mu}_n^N = \sum_{i=1}^{N} w_n^i \delta_{\bar{\btheta}_n^i}
\end{equation}
is obtained and its convergence towards $\bar \mu_n$ has to be established. Finally, after the resampling step, a standard piece of analysis proves the convergence of 
\begin{equation}
	\mu_n^N = \frac{1}{N} \sum_{i=1}^{N} \delta_{\btheta_n^i}
\end{equation}
to $\bar \mu_n$. Below, we provide three lemmas for the conditional convergence of $\check{\mu}_{n-1}^N$, $\tilde{\mu}_n^N$ and $\mu_n^N$, respectively. Then we combine them in order to prove Theorem \ref{thConvergence} by an induction argument.

%
\subsection{Jittering}

In the jittering step, a new cloud of particles $\{ \bar{\btheta}_n^i \}_{i=1}^N$ is generated by propagating the existing samples across the kernels $\kappa_N (\sfd\btheta | \btheta_{n-1}^i)$, $i = 1, \ldots, N$. This step has been analyzed in \cite{Crisan18bernoulli} in the context of the NPF. Several types of kernels can be used. In general, there is a trade-off between the number of particles that are changed using this kernel and the ``amount of perturbation'' that can be applied to each particle. For this reason, we let the jittering kernel $\kappa_N$ depend explicitly on $N$. For our analysis, assumption A.\ref{asKernel} is sufficient.  

The convergence results to be given in this appendix are presented in terms of upper bounds for the $L_p$ norms of the approximation errors. For a random variable $z$, its $L_p$ norm is $\| z \|_p = \mbE\left[ |\boldsymbol{z}|^p \right]^{\frac{1}{p}}$. The approximate measures generated by the nested filter, e.g., $\mu_n^N$, are measured-valued random variables. Therefore, integrals of the form $(h,\mu_n^N)$, for some $h \in B(D)$, are real random variables and it makes sense to evaluate the $L_p$ norm of the random error $(h,\mu_n^N)-(h,\bar \mu_n)$. We start with the approximation $\check \mu_{n-1}^N$ produced after the jittering step at time $n$.   

\begin{Lema} \label{lmSampling}
	Let the sequence of observations $y_{1:n}$ be arbitrary but fixed. If $h \in B(D)$, A.\ref{asKernel} holds and
	\begin{equation}
	\| (h,\mu_{n-1}^{N}) - (h,\bar \mu_{n-1}) \|_p \le \frac{c_{n-1}\|h\|_\infty}{\sqrt{N}}
	\label{eqHypoLemmaKernel}
	\end{equation} 
	for some $p \ge 1$ and a constant $c_{n-1} < \infty$ independent of $N$, then 
	\begin{equation}
	\| (h,\check \mu_{n-1}^{N}) - (h,\bar \mu_{n-1}) \|_p \le \frac{c_{1,n}\|h\|_\infty}{\sqrt{N}},
	\label{eqLemmaKernel}
	\end{equation}
	where the constant $c_{1,n} < \infty$ is also independent of $N$. 
\end{Lema}

\noindent {\bf Proof:} The proof of this Lemma is identical to the proof of \cite[Lemma 3]{Crisan18bernoulli}. $\QED$


%
\subsection{Computation of the weights}

In order to analyze the errors at the weight computation step we rely on assumption A.\ref{asOnG}. An upper bound for the error in the weight computation step is established next.

\begin{Lema}
	Let the sequence of observations $y_{1:n}$ be arbitrary but fixed, choose any $h \in B(D)$ and some $p \ge 1$. If assumptions A.\ref{asLikelihood} and A.\ref{asOnG} hold, and 
	\begin{equation}
	\| 
	(h,\check \mu_{n-1}^{N}) - (h,\bar \mu_{n-1}) \|_p \le \frac{c_{1,n}\|h\|_\infty}{\sqrt{N}}
	\label{eqHypoLemmaWeight-2}
	\end{equation}
	for some constant $c_{1,n} < \infty$ independent of $N$, then
	\begin{equation}
	\| 
	(h,\tilde \mu_{n}^{N}) - (h,\bar \mu_{n}) 
	\|_p \le \frac{c_{2,n}\|h\|_\infty}{\sqrt{N}},
	\label{eqBarMu} 
	\end{equation}
	where the constant $c_{2,n} < \infty$ is independent of $N$. 
	\label{lmWeights}
\end{Lema}

\noindent \textbf{Proof:} We address the characterization of the weights and, therefore, of the approximate measure $\tilde \mu_n^{N} = \sum_{i=1}^N w_n^{i} \delta_{\bar \theta_n^{i}}$. From the definition of the projective product in \eqref{eqDefProjProd}, the integrals of $h$ w.r.t. $\bar \mu_n$ and $\tilde \mu_n^N$ can be written as
\begin{equation}
(h,\bar \mu_n) = \frac{
	(\bar u_n h,\bar \mu_{n-1})
}{
	(\bar u_n, \bar \mu_{n-1})
}, \quad \mbox{and} \quad (h, \tilde \mu_n^{N}) = \frac{
	(\hat u_n h, \check \mu_{n-1}^{N})
}{
	(\hat u_n, \check \mu_{n-1}^{N})
}, \label{eqBayes0}
\end{equation}
respectively. From \eqref{eqBayes0} one can write the difference $(h, \tilde \mu_n^{N}) - (h, \bar \mu_n)$ as 
\begin{equation}
(h, \tilde \mu_n^{N}) - (h, \bar \mu_n) = \frac{
	(h\hat u_n,\check \mu_{n-1}^N) - (h\bar u_n,\bar \mu_{n-1})
}{
	(\bar u_n,\bar \mu_{n-1})
} + (h,\tilde \mu_n^N) \frac{
	(\bar u_n,\bar \mu_{n-1}) - (\hat u_n,\check \mu_{n-1}^N)
}{
	(\bar u_n,\bar \mu_{n-1})
},
\nonumber
\end{equation}
which readily yields the inequality
\begin{equation}
| (h,\tilde \mu_n^{N}) - (h,\bar \mu_{n-1}) | \le \frac{
	| (h \hat u_n, \check \mu_{n-1}^{N}) - (h \bar u_n, \bar \mu_{n-1}) |
}{
	(\bar u_n, \bar \mu_{n-1})
} + \frac{
	\| h \|_\infty | (\hat u_n, \check \mu_{n-1}^{N}) - (\bar u_n,\bar \mu_{n-1}) |  
}{
	(\bar u_n,\bar \mu_{n-1})
}
\label{eqL1Error-0}
\end{equation}
by simply noting that $|(h,\tilde \mu_n^N)| \le \| h \|_\infty$, since $\tilde \mu_n^N$ is a probability measure. From \eqref{eqL1Error-0} and Minkowski's inequality we easily obtain the bound
\begin{eqnarray}
\| (h,\tilde \mu_n^{N}) - (h, \bar \mu_{n-1}) \|_p &\le& \frac{
	1
}{
	(\bar u_n, \bar \mu_{n-1})
} \left[
\| h \|_\infty \| (\hat u_n, \check \mu_{n-1}^{N}) - (\bar u_n, \bar \mu_{n-1}) \|_p 
\right. \nonumber\\
&& \left.
+ \| (h \hat u_n, \check \mu_{n-1}^{N}) - (h \bar u_n, \bar \mu_{n-1}) \|_p,
\right]
\label{eqL1Error-1}
\end{eqnarray}
where $(\bar u_n,\bar \mu_{n-1})>0$ from assumption A.\ref{asOnG}.2.

We need to find upper bounds for the two terms on the right hand side of \eqref{eqL1Error-1}. Consider first the term $\| (h\hat u_n,\check \mu_{n-1}^{N}) - (h \bar u_n,\bar \mu_{n-1}) \|_p$. A simple triangle inequality yields
\begin{equation}
\| (h \hat u_n,\check \mu_{n-1}^{N}) - (h \bar u_n,\bar \mu_{n-1}) \|_p \le 
\| (h \hat u_n,\check \mu_{n-1}^{N}) - (h \bar u_n,\check \mu_{n-1}^{N}) \|_p 
+ \| (h \bar u_n,\check \mu_{n-1}^{N}) - (h \bar u_t,\bar \mu_{n-1}) \|_p.
\label{eqWeights-triangle}
\end{equation}
On one hand, since $\sup_{\btheta \in D} | h(\btheta) \bar u_n(\btheta) | \le \| h \|_\infty \| \bar u_n \|_\infty < \infty$ (see A.\ref{asOnG}.1), it follows from the assumption in Eq. \eqref{eqHypoLemmaWeight-2} that
\begin{equation}
\| (h \bar u_n, \check \mu_{n-1}^{N}) - (h \bar u_n, \bar \mu_{n-1}) \|_p \le \frac{
	c_{1,n}\|h\|_\infty \| \bar u_n \|_\infty
}{
	\sqrt{N}
},
\label{eqBoundSecondTermWeight}
\end{equation}
where $c_{1,n} < \infty$ is a constant independent of $N$. 

On the other hand, we may note that
\begin{equation}
| (h \hat u_n, \check \mu_{n-1}^{N}) - (h \bar u_n,\check \mu_{n-1}^{N}) |^p = \left|
	\frac{1}{N} \sum_{i=1}^N \left(
		h(\bar{\btheta}_n^i) \hat u_n(\bar{\btheta}_n^{i}) - h(\bar{\btheta}_n^i) \bar u_n(\bar{\btheta}_n^{i})
	\right)
\right|^p.
\label{eqWeightsNearTheEnd}
\end{equation}
Let $\mathcal{G}_n$ be the $\sigma$-algebra generated by the random particles $\{ \bar{\btheta}_{1:n-1}^i, \btheta_{0:n-1}^i \}_{1\le i\le N}$ and assume that $p$ is even. Then we can apply conditional expectations on both sides of  \eqref{eqWeightsNearTheEnd} to obtain
\begin{eqnarray}
\mbE\left[
	\left|
		(h \hat u_n, \check \mu_{n-1}^{N}) - (h \bar u_n, \check \mu_{n-1}^{N}) 
	\right|^p \Big| \mathcal{G}_n
\right]  &=& \mbE\left[
	\left(
		\frac{1}{N} \sum_{i=1}^{N} h(\bar{\btheta}_n^i) {m}_n(\bar{\btheta}_n^i)
	\right)^p \Big| \mathcal{G}_n
\right] 
\nonumber
\end{eqnarray}
where the expression on the right hand side has been simplified by using the assumption $\hat u_n(\btheta) = \bar u_n(\btheta) + m_n(\btheta)$ in A.\ref{asLikelihood}. Also from assumption A.\ref{asLikelihood}, the random variables $m_n(\bar{\btheta}_n^i)$ are conditionally independent (given $\mathcal{G}_n$), have zero mean and finite moments of order $p$, $\mbE[ {m}_n(\bar{\btheta}_n^i)^p ] \le \sigma^p < \infty$. If we realise that 
$$
\mbE[ h(\bar{\btheta}_n^i) m_n(\bar{\btheta}_n^i) | \mathcal{G}_n ] = h(\bar{\btheta}_n^i) \mbE[  m_n(\bar{\btheta}_n^i) | \mathcal{G}_n ] = 0
$$ 
and bear in mind the conditional independence of the $m_n(\bar{\btheta}_n^i)$'s, then it is an exercise in combinatorics to show that the number of non-zero terms in
\begin{equation}
	\mbE\left[
		\left(
			\frac{1}{N} \sum_{i=1}^{N} h(\bar{\btheta}_n^i) {m}_n(\bar{\btheta}_n^i)
		\right)^p \Big| \mathcal{G}_n
	\right]  =
	\sum_{i_1} \ldots \sum_{i_p} \mbE\left[
		h(\bar{\btheta}_n^{i_1}) m_n(\bar{\btheta}_n^{i_1}) \ldots h(\bar{\btheta}_n^{i_p}) m_n(\bar{\btheta}_n^{i_p}) 
	\Big| \mathcal{G}_n 
	\right]
	\nonumber
\end{equation}
is at most $\tilde c^p N^{\frac{p}{2}}$, for some constant $\tilde c^p<\infty$ independent of $N$ and $h$. Since each of the non-zero terms is upper bounded by $\mbE\left[ (h(\bar{\btheta}_n^i) m_n(\bar{\btheta}_n^i))^p \Big| \mathcal{G}_n \right] \le \| h \|_{\infty}^p \sigma^p < \infty$ (using A.\ref{asLikelihood} again), then it follows that 
\begin{equation}
	\mbE\left[
		\left|
			(h \hat u_n, \check \mu_{n-1}^{N}) - (h \bar u_n, \check \mu_{n-1}^{N}) 
		\right|^p
	\right]  = 
	\mbE\left[
		\left(
			\frac{1}{N} \sum_{i=1}^{N} h(\bar{\btheta}_n^i) {m}_n(\bar{\btheta}_n^i)
		\right)^p \Big| \mathcal{G}_n
	\right]  \le 
	\frac{\tilde{c}^p \sigma^p \| h \|_{\infty}^p }{N^{\frac{p}{2}}}
	\label{eqBoundFirstTermWeight}
\end{equation}
for even $p$. Given \eqref{eqBoundFirstTermWeight}, it is straightforward to show that the same result holds for every $p \ge 1$ using Jensen's inequality. Finally, since the bound on the right hand side of \eqref{eqBoundFirstTermWeight} is independent of $\mathcal{G}_n$, we can take expectations on both sides of the inequality and obtain that
\begin{equation}
\| (h \hat u_n, \check \mu_{n-1}^{N}) - (h \bar u_n,\check \mu_{n-1}^{N}) \|_p \le \frac{
	\tilde{c} \sigma \| h \|_\infty
}{
	\sqrt{N}
} .
\label{eqBoundFirstTermWeight_bis}
\end{equation}
Substituting \eqref{eqBoundFirstTermWeight_bis} and \eqref{eqBoundSecondTermWeight} into \eqref{eqWeights-triangle} yields
\begin{equation}
\| (h \hat u_n, \check \mu_{n-1}^{N}) - (h \bar u_n,\bar \mu_{n-1}) \|_p \le \frac{
	c_n' \| h \|_{\infty}^p  \| \bar u_n \|_\infty
}{
	\sqrt{N}
} ,
\label{eqBoundTerm_u}
\end{equation}
where $c_n' = c_{1,n} + \tilde{c} \sigma$ is a constant independent of $N$.

The same argument leading to the bound in \eqref{eqBoundTerm_u} can be repeated, step by step, on the norm $\| (\hat u_n,\check \mu_{n-1}^N) - (\bar u_n,\bar \mu_{n-1}) \|_p$ (simply taking $h(\btheta) = 1$), to arrive at
\begin{equation}
\| (\hat u_n, \check \mu_{n-1}^{N}) - (\bar u_n, \bar \mu_{n-1}) \|_p \le \frac{
	c_n' \|\bar u_n\|_\infty
}{
	\sqrt{N}
}.
\label{eqBoundTerm_hu}
\end{equation}

To complete the proof, we substitute \eqref{eqBoundTerm_u} and \eqref{eqBoundTerm_hu} back into \eqref{eqL1Error-1} and so obtain
\begin{equation}
\| (h,\tilde \mu_n^{N}) - (h, \bar \mu_{n-1}) \|_p \le \frac{
	c_{2,n}\| h \|_\infty
}{
	\sqrt{N}
}, \nonumber
\end{equation}
where the constant 
$
c_{2,n} = \| \bar u_n \|_\infty \left(2 c_n' \right) / (\bar u_t,\bar \mu_{t-1}) < \infty
$ is independent of $N$. $\QED$

\subsection{Resampling}

The quantification of the error in the resampling step of the nested filter is a standard piece of analysis, well known from the particle filtering literature (see, e.g., \cite{Bain08}). We can state the following result.

\begin{Lema} \label{lmResampling}
	Let the sequence of observations $y_{1:n}$ be arbitrary but fixed. If $h \in B(D)$ and
	\begin{equation}
	\| 
	(h,\tilde \mu_{n}^{N}) - (h, \bar \mu_{n})
	\|_p \le \frac{c_{2,n}\| h \|_\infty}{\sqrt{N}}
	\label{eqAssResampling-theta}
	\end{equation}
	for a constant $c_{2,n} < \infty$ independent of $N$, then 
	$$
	\| 
	(h,\mu_{n}^{N}) - (h,\bar \mu_{n}) 
	\|_p \le \frac{c_{3,n}\| h \|_\infty}{\sqrt{N}},
	$$
	where the constant $c_{3,n} < \infty$ is independent of $N$ as well. 
\end{Lema}
\noindent {\bf Proof:} See, e.g., the proof of \cite[Lemma 1]{Miguez13b}. $\QED$

%
\subsection{An induction proof for Theorem \ref{thConvergence}}

Finally, we can put Lemmas \ref{lmSampling}, \ref{lmWeights} and \ref{lmResampling} together in order to prove the inequality \eqref{eqBoundResampling}  by induction in $n$. At time $n=0$, we draw $\btheta_0^i$, $i=1,\ldots,N$, independently from the prior $\mu_0=\bar \mu_0$ and it is straightforward to show that $\| (h, \mu_{0}^N) - (h, \bar \mu_{0}) \|_p \le \frac{ c_0 \| h \|_{\infty}	}{\sqrt{N}}$, where $c_0$ does not depend on $N$.
	
Assume that, at time $n-1$, 
	\begin{equation}
		\nonumber
		\| (h, \mu_{n-1}^N) - (h, \bar \mu_{n-1}) \|_p \le \frac{
			c_{n-1} \| h \|_{\infty}
		}{
			\sqrt{N}
		}
	\end{equation}
where $c_{n-1} < \infty$ is independent of $N$. Then, we simply apply Lemmas \ref{lmSampling}, \ref{lmWeights} and \ref{lmResampling} in sequence to obtain
	\begin{equation}
	\nonumber
	\| (h, \mu_{n}^N) - (h, \bar \mu_{n}) \|_p \le \frac{
		c_{n} \| h \|_{\infty}
	}{
		\sqrt{N}
	}
	\end{equation}
for a constant $c_n = c_{3,n} < \infty$ independent of $N$. $\QED$


\section{Simplification of the inverse $(\boldsymbol{S}^i)^{-1}$} \label{sInverseSimpl}

The predictive covariance of the observation vector $\by_n$ is a $d_y \times d_y$ matrix $\boldsymbol{S}_n$. Inverting $\boldsymbol{S}_n$ has a cost $\mathcal{O}(d_y^3)$, which can become intractable. Assuming that variables located ``far away" in the circumference of the Lorenz 96 model have small correlation we can approximate $\boldsymbol{S}_n$ as a block diagonal matrix, namely, $\hat{\boldsymbol{S}}_n = \boldsymbol{S}_n \odot \boldsymbol{M}$, where $\odot$ denotes element-wise product,
\begin{equation} \label{eqMask}
\boldsymbol{M}=
\begin{bmatrix}
\boldsymbol{\mathrm{1}} & \boldsymbol{\mathrm{0}} &  \ldots & \boldsymbol{\mathrm{0}} \\
\boldsymbol{\mathrm{0}} & \boldsymbol{\mathrm{1}} &  \ldots &  \boldsymbol{\mathrm{0}} \\
\vdots & \vdots &  \ddots  &\vdots \\
\boldsymbol{\mathrm{0}} & \boldsymbol{\mathrm{0}} &\ldots &  \boldsymbol{\mathrm{1}} \\
\end{bmatrix}
\end{equation}
is a mask matrix and $\boldsymbol{\mathrm{0}}$ and $\boldsymbol{\mathrm{1}}$ are, respectively, matrices of zeros and ones of dimension $d_q \times d_q$. There are $Q$ blocks in the diagonal of $\boldsymbol{M}$, hence $d_y = Qd_q$. The original matrix could contain some non-zero values where the zero blocks of $\boldsymbol{M}$ are placed, however their values are assumed close to zero. The resulting matrix,
$$
\hat{\boldsymbol{S}}=\begin{bmatrix}
\bar{\boldsymbol{S}}_1 & \boldsymbol{\mathrm{0}} &  \ldots &  \boldsymbol{\mathrm{0}} \\
\boldsymbol{\mathrm{0}} & \bar{\boldsymbol{S}}_2 &  \ldots &  \boldsymbol{\mathrm{0}} \\
\vdots & \vdots &  \ddots &\vdots \\
\boldsymbol{\mathrm{0}} & \boldsymbol{\mathrm{0}} & \ldots &  \bar{\boldsymbol{S}}_{Q} \\
\end{bmatrix},
\quad \mbox{is easily inverted as}
\quad
\hat{ \boldsymbol{S}}_n^{-1}=
\begin{bmatrix}
\bar{\boldsymbol{S}}_1^{-1} & \boldsymbol{\mathrm{0}} &  \ldots &  \boldsymbol{\mathrm{0}} \\
\boldsymbol{\mathrm{0}} & \bar{\boldsymbol{S}}_2^{-1} &  \ldots &  \boldsymbol{\mathrm{0}} \\
\vdots & \vdots &  \ddots &\vdots \\
\boldsymbol{\mathrm{0}} & \boldsymbol{\mathrm{0}} & \ldots &  \bar{\boldsymbol{S}}_{Q}^{-1} \\
\end{bmatrix}
$$
with a computational cost $\mathcal{O}(Qd_q^3) = \mathcal{O}(\frac{d_y^3}{Q^2})$.

\bibliographystyle{plain} 
\bibliography{bibliografia,biblio_ines} 

\end{document}